\documentclass[aps,amsmath,amssymb,showpacs,showkeys,longbibliography]{revtex4-2}
\usepackage[dvips]{graphicx}
\usepackage{times}
\usepackage{braket}
\usepackage{xcolor}
\usepackage{orcidlink}
\usepackage{hyperref}
\hypersetup{
  colorlinks=true,
  urlcolor=magenta,
  linkcolor=red,
  citecolor=blue
}
\usepackage{float}
\usepackage{multirow}
\usepackage{mathptmx}
\usepackage[sort&compress]{natbib}
\usepackage[shortcuts]{extdash}

\begin{document}
\title{Black Hole Quasi-Periodic Oscillations in the Presence of Gauss–Bonnet 
Trace Anomaly}

\author{Rupam Jyoti Borah\orcidlink{0009-0005-5134-0421}}
\email[Email: ]{rupamjyotiborah856@gmail.com}

\author{Umananda Dev Goswami\orcidlink{0000-0003-0012-7549}}
\email[Email: ]{umananda@dibru.ac.in}

\affiliation{Department of Physics, Dibrugarh University, Dibrugarh 786004, 
Assam, India}

\begin{abstract}
We investigate the effects of the Gauss–Bonnet (GB) gravitational trace 
anomaly on the circular motion of test particles around black holes (BHs) and 
its implications for quasi-periodic oscillations (QPOs) in various theoretical 
models. Beginning with the equations of motion, we study the effects on 
effective potential, angular momentum, specific energy, and the innermost 
stable circular orbit (ISCO) induced by the anomaly parameter $\alpha$. The 
fundamental frequencies are calculated. Moreover, we examine several QPO 
models, including PR, RP, WD, TD, and ER2–ER4, and study the relationship 
between the upper and lower QPO frequencies as well as the corresponding 
resonance radii for frequency ratios of 1:1, 3:2, 4:3, and 5:4. Our results 
show that increasing $\alpha$ leads to deviations from the 
Schwarzschild case in both upper and lower QPO frequencies correlations and 
QPO orbital radii, with model-dependent trends. Further, we constrain the BH 
parameters using the observational data using MCMC analysis. Finally, 
we calculate the upper and lower QPO frequencies for a few BH candidates on 
the basis of the RP model using the constrained parameter values and find a 
good agreement with the observed results.

\end{abstract}


\maketitle                                                                      

\section{Introduction}
General Relativity (GR) has been the most successful and extensively 
tested theory of gravity. Its predictions have been tested to a high precision 
through a wide range of experiments and astronomical observations. Among 
these, the recent detection of gravitational waves (GWs) by the LIGO 
collaboration \cite{L1,L2,L3,L4,L4} and the capturing of the image of black
holes (BHs) by the Event Horizon Telescope (EHT) \cite{E1,E2,E3,E4,E5,E6} are 
the major achievements in gravitational physics as well as in observational 
astrophysics. Despite these empirical successes, GR faces several profound 
theoretical challenges. Notably, it does not provide a satisfactory 
explanation for the accelerated expansion of the Universe, and it fails to 
explain the flat galaxy rotation curves obtained from observations, which are 
typically associated with dark energy and dark matter riddles, respectively 
\cite{F1,F2,F3,F4}. Furthermore, GR is a classical theory and fails to 
incorporate quantum effects that are expected to dominate at high energy 
scales. In contrast, the Standard Model (SM) of particle physics 
\cite{SM1,SM2} is a very successful quantum theory of matters-radiation and 
their interactions which unifies the electromagnetic, weak, and strong 
interactions within the framework of Quantum Field Theory (QFT) 
\cite{QFT1,QFT2}. However, gravity remains excluded from this unification, 
primarily because GR is non-renormalizable and thus incompatible with the 
quantization procedures employed in QFT. 

The unification of GR with the SM remains a central challenge in modern 
theoretical physics and continues to be an active area of research. In 
response to this challenge, several promising frameworks of Quantum Gravity 
(QG) have been developed to describe gravity within the framework of quantum 
physics. Among these most extensively studied approaches are Loop Quantum 
Gravity (LQG) \cite{LQG1,LQG2,LQG3,LQG4} and String Theory 
\cite{ST1,ST2,ST3,ST4,ST5}. LQG is developed within the framework of the 
Ashtekar formalism \cite{Asthe1,Asthe2} of GR, and it predicts that spacetime becomes 
discrete at extremely high energy scales. On the other hand, String Theory 
adopts a fundamentally different approach, proposing the existence of a boson 
particle known as the graviton, which acts as the mediator of the 
gravitational interaction.

Notwithstanding being the most extensively studied candidates for a theory of 
QG, both LQG and String Theory lack experimental validation. This problem 
arises because, to test the predictions of these theories, we require a very 
large amount of energy which is not currently achievable by existing 
experimental technologies. As a result, the direct empirical validation of 
these theories remains infeasible. Consequently, researchers have begun 
exploring alternative approaches to probe possible QG effects within more 
accessible energy regimes. One promising alternative is the Effective Field 
Theory (EFT) approach. EFT offers a framework to study the QG effects within 
an experimentally accessible energy scale, avoiding the need for ultra-high 
energies. In this formulation, GR is treated as a low-energy effective theory, 
valid up to a certain cutoff scale. This allows for the systematic inclusion 
of quantum corrections while preserving the agreement with classical GR in 
the low-energy regime. Numerous studies have been done in this approach, 
employing GR as an EFT to explore quantum gravitational phenomena 
(e.g.~see \cite{GREFT1,GREFT2,GREFT3,GREFT4,GREFT5,GREFT6,GREFT7}). 

BHs are regions of spacetime where the gravitational field becomes so strong 
that not even light can escape. They represent one of the most fundamental and 
profound predictions of GR. Thus near a BH, gravity becomes extremely strong, 
and as a result, the small quantum effects arising from EFT corrections can 
become significant. The signatures of such effects can be observed in different 
observables of BHs. Different properties of BHs, such as BH thermodynamics, 
quasinormal modes (QNMs), and BH shadows, have been extensively studied in 
this direction in the literature \cite{thermo1,thermo2,QNM3,QNM4,QNM5,Sha1,
Sha2,Sha3}.

One of the significant quantum effects that naturally emerges in the EFT 
framework is the trace anomaly, also known as the conformal anomaly \cite{Tra1,Tra2,Tra3}. 
The classical field theories are conformally invariant and they exhibit a 
vanishing trace of the energy-momentum tensor. This symmetry is generally 
broken at the quantum level. In curved spacetime, this anomaly leads to a 
non-zero expectation value of the trace of the energy-momentum tensor. The 
trace anomaly captures the effects of quantum fields on the curvature of 
spacetime by adding specific geometric terms to the effective action. These 
terms depend on how spacetime is curved and include quantities like the 
square of the Weyl tensor, which measures the tidal distortion of spacetime, 
and the Gauss-Bonnet (GB) invariant, which combines different curvature 
components 
into a topological quantity. Together, these curvature dependent terms reflect 
how quantum corrections modify the classical description of gravity. These 
anomaly-induced corrections have profound implications for the semiclassical 
dynamics of spacetime, influencing phenomena like BH QNMs, \cite{qnm1,qnm2}, 
quasi-periodic oscillations (QPOs) of BH \cite{qp1,qp2,qp3,qp4,qp5,qp6,qp7}, BH 
evaporation \cite{Evap1,Evap2,Evap3}, quantum backreaction \cite{Bk1,Bk2}, 
and early universe cosmology 
\cite{Cosmo1,Cosmo2,Cosmo3} etc. Within the EFT framework, the trace anomaly 
provides a concrete and calculable imprint of quantum effects in gravity, 
enabling theoretical predictions that remain compatible with the classical 
limit of GR. The effects of quantum corrections from EFT on gravity are 
typically very weak. However, they become significant in regions of extreme 
gravity, such as near the event horizon of a BH as mentioned earlier.

BH's QPOs are characteristic features observed in the X-ray light curves of 
accreting BH systems, especially in X-ray binaries. These oscillations appear 
as narrow peaks in the power density spectrum, indicating the presence of 
periodic or near-periodic modulations in the X-ray flux. QPOs are classified 
into two categories: low-frequency QPOs (LFQPOs) \cite{lqpo1,lqpo2,lqpo3} and 
high-frequency QPOs (HFQPOs) \cite{hqpo1,hqpo2}, based on their characteristic 
frequencies. QPOs are thought to originate from the innermost regions of the 
accretion disk, close to the event horizon, where strong gravity effects are 
significant. These frequencies are thought to correspond to fundamental modes 
of motion and are sensitive to the spacetime geometry surrounding the BH. 
Following their initial discovery via spectral and timing analyses in X-ray 
binaries \cite{xr}, QPOs have been extensively investigated from both 
observational and theoretical perspectives 
\cite{obs1,obs2,obs3,QPO1,QPO2,QPO3}. These efforts have led to the 
development of many theoretical works aiming to interpret the origin of QPOs 
within the framework of GR and accretion physics. A variety of models have 
been developed to capture the complex interplay between spacetime curvature, 
disk dynamics, and radiative processes in the vicinity of BHs 
\cite{QPO11,QPO12,QPO13,QPO14}. These theoretical models aim to reproduce the 
observed QPO properties and their links to spectral states, which helps in 
the better understanding of the nature of the strong gravitational field near 
the BH. In spite of a large number of studies, the exact physical mechanisms 
responsible for generating QPOs remain an open question, with several 
competing models attempting to explain their origin. Continued advancements 
in observational techniques and theoretical modeling are essential to uncover 
the underlying physics driving these intriguing phenomena. 

Thus, this study focuses on the effects of an EFT, which is based on the 
gravitational trace anomaly in GR, on the circular motion of a test particle 
around a BH and on the QPOs of the corresponding BH, and then constrains the 
parameters of the underlying theory using the QPO data. In Section~\ref{2}, we 
briefly discuss the BH solution in the presence of gravitational trace anomaly 
in GR. In Section~\ref{3}, we discuss the dynamics of a test particle around 
the BH in the said theory. In Section~\ref{4}, we analyze the quasi-periodic 
and harmonic oscillations through the fundamental frequencies in this 
theoretical framework. The QPO models and QPO orbits for the BH influenced by 
trace anomaly have been studied in Section~\ref{5}. In Section~\ref{6}, 
constraining the BH parameters using QPO data and MCMC analysis has been 
made. Finally, in Section~\ref{7}, we summarize and conclude our work.

\section{Black Hole Solutions in Trace Anomaly-Induced Gravity} \label{2}
As BHs are regions of spacetime where gravitational effects become extremely 
strong, studying BHs influenced by the gravitational trace anomaly therefore 
offers a powerful avenue for exploring the physical implications of quantum 
behaviours in curved spacetime. The gravitational theory with the trace 
anomaly was first introduced in Ref.~\cite{Ga1} and the BHs incorporating the 
gravitational trace anomaly were first studied in Ref.~\cite{Tsu1}. This 
section provides a brief overview of the main steps involved in deriving the 
BH solution with the gravitational trace anomaly. 

The gravitational action that avoids the problem of strong coupling in GR
and is suitable for the quantum level is given by (see Refs.~\cite{Tsu1,Ga1})
\begin{equation}
S = S_{R\bar{R}} + S_{A}. \label{eq1}
\end{equation} 
In this action \eqref{eq1} the first term $S_{R\bar{R}}$ is given as
\begin{equation}
S_{R\bar{R}} = M^2\!\! \int\! d^{\,4}x\sqrt{-g}\,R - \bar{M}^2\!\!\int\! d^{\,4}x\sqrt{-\bar{g}}\,\bar{R}, \label{eq2}
\end{equation}
where $g$ is the determinant of the metric tensor $g_{\mu\nu}$ and $\bar{g}$ 
is the determinant of $\bar{g}_{\mu\nu}$. The metric tensor $\bar{g}_{\mu\nu}$ 
is related to $g_{\mu\nu}$ by the conformal transformation:
\begin{equation}
\bar{g}_{\mu\nu} = e^{-2\phi}g_{\mu\nu} \label{eq3}
\end{equation}  
with $\phi$ identified as a massless scalar degree of freedom of the conformal 
field. $R$ and $\bar{R}$ are Ricci scalars evaluated in terms of $g_{\mu\nu}$ 
and $\bar{g}_{\mu\nu}$ respectively. $M$ is a constant that is related to 
reduced Planck mass $M_{Pl}$ as $M = M_{pl}/ \sqrt{2}$. $\bar{M}$ is a new 
mass scale which is significantly smaller than $M$. The trace anomaly action 
$S_A$ is expressed as \cite{Tsu1}
\begin{equation}
S_A[\bar{g}, \phi] = \int\! d^{\,4}x\, \sqrt{-\bar{g}} \left[ 
  -\alpha \left( \phi\, \bar{\mathcal{G}} 
  - 4\, \bar{G}^{\mu\nu} \bar{\nabla}_{\mu} \phi\, \bar{\nabla}_{\nu} \phi 
  + 8\, \bar{X} \bar{\Box} \phi - 8\, \bar{X}^2 \right) 
  + \beta \phi \bar{W}^2 
\right], \label{eq4}
\end{equation} 
where $\alpha$ and $\beta$ are the coupling parameters, $\bar{G}^{\mu\nu}$~is
the conformally transformed Einstein tensor, $\bar{\mathcal{G}}$ and 
$\bar{W}^2$ respectively are the conformally transformed Gauss-Bonnet and 
square of Weyl terms, whose forms are $\bar{\mathcal{G}} = \bar{R}^2-
4\,\bar{R}_{\mu\nu} \bar{R}^{\mu\nu} + 
\bar{R}_{\mu\nu\rho\sigma}\bar{R}^{\mu\nu\rho\sigma}$ and 
$W^2= \bar{R}^2/2 - 2\,\bar{R}_{\mu\nu}\bar{R}^{\mu\nu}+
\bar{R}_{\mu\nu\rho\sigma}\bar{R}^{\mu\nu\rho\sigma}$ respectively. 
Further, $\bar{X}$ can be express as $\bar{X} = 
-(1/2)\,\bar{g}^{\mu\nu}{\bar{\nabla}_{\mu}}{\bar{\nabla_{\nu}}}\phi$.
Moreover, all the quantities with a bar in Eq.~\eqref{eq4} are in the 
conformally transformed frame. Thus, the total action $S$ in the conformally 
transformed frame can be represented as follows:
\begin{equation}
S = \int \mathrm{d}^4x \sqrt{-\bar{g}} \left[ M^2 e^{2\phi} \left( \bar{R} - 12 \bar{X} \right) 
- \bar{M}^2 \bar{R} 
- \alpha \left( \phi \bar{\mathcal{G}} - 4\, \bar{G}^{\mu\nu} \bar{\nabla}_{\mu} \phi\, \bar{\nabla}_{\nu} \phi 
+ 8\, \bar{X} \bar{\Box} \phi - 8\, \bar{X}^2 \right) 
+ \beta \phi \bar{W}^2 \right].
\end{equation}
The action with $\beta = 0$ i.e.\ action with only GB trace anomaly term is
\begin{equation}
S = \int \mathrm{d}^4x \sqrt{-\bar{g}} \left[ M^2 e^{2\phi} \left( \bar{R} - 12 \bar{X} \right) 
- \bar{M}^2 \bar{R} 
- \alpha \left( \phi \bar{\mathcal{G}} - 4\, \bar{G}^{\mu\nu} \bar{\nabla}_{\mu} \phi\, \bar{\nabla}_{\nu} \phi 
+ 8 \,\bar{X} \bar{\Box} \phi - 8 \bar{X}^2 \right) \right].
\end{equation}
In the conformally transformed frame with metric tensor $\bar{g}_{\mu\nu}$, if
we consider a static and spherically symmetric spacetime, the line element can 
be expressed as
\begin{equation}
d{\bar{s}}^{\,2} = -\bar{f}(\bar{r})\,dt^2 + \frac{1}{\bar{h}\,(\bar{r})}d\bar{r}^{\,2} + {\bar{r}}^{\,2}(d{\theta}^2 +\sin^{2}{\theta}d{\varphi}^2), \label{eq7}
\end{equation}
where $\bar{f}(\bar{r})$ and $\bar{h}(\bar{r})$ are the metric functions
in the conformal frame. Also, it is assumed that the scalar field $\phi$ is 
a function of radial coordinate $\bar{r}$ only, i.e.~$\phi = \phi(\bar{r})$. 
The line element associated with the original frame with metric tensor 
$g_{\mu\nu} = e^{2\phi}\bar{g}_{\mu\nu}$ is
\begin{equation}
ds^2 = -f(r)\,dt^2 + \frac{1}{h(r)}\,dr^2 + r^2(d{\theta}^2 + sin^2{\theta}d{\varphi}^2). \label{eq8}
\end{equation}
These two line elements are related by the transformation 
$d{\bar{s}}^2 = e^{-2\phi} ds^2$ and hence the temporal, radial and angular 
components of these two line elements are related as
\begin{equation}
f(r) = \bar{f}(\bar{r})\, e^{2\phi}, \;\;\; 
h(r) = \bar{h}(\bar{r}) \left[ 1 + \bar{r} \phi'(\bar{r}) \right]^2, \;\;\; 
r = \bar{r} e^{\phi}, \label{eq9}
\end{equation}
where $\phi^{\prime}(\bar{r})$ is the derivative of $\phi(\bar{r})$ with 
respect to $\bar{r}$.

Upon evaluating the $tt$ and $\bar{r}\bar{r}$ components of the field 
equations, which are obtained by the variation of action \eqref{eq1}, we 
derive the following differential equations for 
$\bar{f}(\bar{r})$ and $\bar{h}(\bar{r})$ \cite{Tsu1}:
\begin{align}
\frac{\bar{f}'}{\bar{f}} & = \frac{\bar{M}^2 (\bar{h} - 1) - M^2 e^{2\phi} \left[ \bar{h} (3\bar{r} \phi' + 1)(\bar{r} \phi' + 1) - 1 \right] - \alpha \bar{h} \phi'^2 \left[ \bar{h} \left\{ \bar{r} \phi' (3\bar{r} \phi' + 8) + 6 \right\} - 2 \right]}{\bar{h} \left[ M^2 \bar{r} (\bar{r} \phi' + 1) e^{2\phi} + 2\alpha \phi' \left( \bar{h} \left\{ 3 + \bar{r} \phi' (\bar{r} \phi' + 3) \right\} - 1 \right) - \bar{M}^2 \bar{r} \right]}, \label{eq10}\\[8pt]
\frac{\bar{h}'}{\bar{h}} & - \frac{\bar{f}'}{\bar{f}} = 
\frac{2(\phi'^2 - \phi'') \left[ M^2 \bar{r}^2 e^{2\phi} + 2\alpha \left( \bar{r}^2 \bar{h} \phi'^2 + 2 \bar{r} \bar{h} \phi' + \bar{h} - 1 \right) \right]}
{\bar{r} M^2  e^{2\phi} (\bar{r} \phi' + 1) - \bar{r} \bar{M}^2 + 2\alpha \phi' \left( \bar{r}^2 \bar{h} \phi'^2 + 3 \bar{r} \bar{h} \phi' + 3 \bar{h} - 1 \right)}. \label{eq11}
\end{align}
Again, the variation of the action \eqref{eq1} with respect to $\phi$ in the 
frame with metric $g_{\mu\nu}$ yields \cite{Tsu1}:
\begin{equation}
2 M^2 R - \alpha\,\mathcal{G} + \beta W^2 = 0. \label{12}
\end{equation}
In the conformally transformed frame with metric ${\bar{g}}_{\mu\nu}$, 
Eq.~\eqref{12} can be expressed as
\begin{equation}
2 \bar{M}^2 \bar{R} - \alpha\,\bar{\mathcal{G}} + \beta \bar{W}^2 = 0. \label{eq13}
\end{equation} 
From Eq.~\eqref{eq13} we obtain,
\begin{align}
\big[2 \bar{r}^2 \bar{f} \bar{h} \bar{f}'' 
& - \bar{r} \bar{f}' \big( \bar{r} \bar{h} \bar{f}' - \bar{r} \bar{f} \bar{h}' - 4 \bar{f} \bar{h} \big) 
+ 4 \bar{f}^2 (\bar{h} - 1) + 4\, \bar{r} \bar{f}^2 \bar{h}' \big] \bar{M}^2
\notag\\[5pt] & + 2\alpha \big[ 2 \bar{f} \bar{h} (\bar{h} - 1) \bar{f}'' 
 - \bar{f}' \left( \bar{h}^2 \bar{f}' - 3 \bar{f} \bar{h} \bar{h}' - \bar{f}' \bar{h} + \bar{f} \bar{h}' \right) \big] = 0. \label{eq14}
\end{align}
The general forms of the solutions for $\bar{f}(\bar{r})$, $\bar{h}(\bar{r})$ 
and $\bar{\phi}(\bar{r})$ under the Schwarzschild ansatz can be expressed as 
follows:
\begin{equation}
f(\bar{r}) = \left(1 - \frac{\bar{r}_h}{\bar{r}}\right) \left[ 1 + \sum_{i=1}^{\infty} \bar{f}_i(\bar{r}) \alpha^i \right]\!,\;\; 
h(\bar{r}) = \left(1 - \frac{\bar{r}_h}{\bar{r}}\right) \left[ 1 + \sum_{i=1}^{\infty} \bar{h}_i(\bar{r}) \alpha^i \right]\!,\;\; 
\phi(\bar{r}) = \phi_0 + \sum_{i=1}^{\infty} \phi_i(\bar{r}) \alpha^i. 
\label{eq15}
\end{equation}
Here, $\bar{f}_i(\bar{r})$, $\bar{h}_i(\bar{r})$ and $\phi_i(\bar{r})$ 
denote functions of the radial coordinate $\bar{r}$, and $\bar{r}_h$ 
represents the event horizon radius of the BH in the barred frame. However, for 
consistency with the subsequent analysis performed in the unbarred coordinate 
system defined by $g_{\mu\nu}$, i.e., in the unbarred frame, we identify the 
physical event-horizon radius as $r_h = 2 \mathcal{M}$, with $\mathcal{M}$ 
being the mass of the BH. For validity of Eq.~\eqref{eq15} requires the 
following condition to be fulfilled:
\begin{equation}
\frac{\alpha}{{\bar{M}}^2 {{\bar{r}_h}^2}} << 1. \label{eq16}
\end{equation}
Substituting Eq.~\eqref{eq15} into Eqs.~\eqref{eq10}, \eqref{eq11} and 
\eqref{eq14}, and solving the resulting differential equations, the BH 
solutions with scalar hair can be systematically derived at each order in 
$\alpha$. The expressions for $\bar{f}(\bar{r})$, $\bar{h}(\bar{r})$ and 
$\phi(\bar{r})$ up to the order of $\alpha^2$ are given below \cite{Tsu1}:
\begin{align}
\bar{f}(\bar{r}) = &\left(1 - \frac{\bar{r}_h}{\bar{r}}\right) \biggl[ 
1 + \frac{23\, \bar{r}^{\,2} + 11 \bar{r}_h \bar{r} + 6\, \bar{r}_h^{\,2}}{18 \bar{M}^2 \bar{r}_h \bar{r}^3}\, \alpha - \frac{1}{32400\, M^{\,2} \bar{M}^{\,4} \bar{r}_h^{\,3} \bar{r}^{\,6}}\nonumber \\ 
& \times \Bigl\{\bar{M}^{\,2} \left(38731 \bar{r}^{\,5} + 26371 \bar{r}_h \bar{r}^{\,4} + 22721 \bar{r}_h^{\,2} \bar{r}^{\,3} 
- 769\, \bar{r}_h^{\,3} \bar{r}^{\,2} - 5572\, \bar{r}_h^{\,4} \bar{r} - 9300\, \bar{r}_h^{\,5}\right) e^{-2 \phi_0}\nonumber \\
& - M^{\,2} \left(6979\, \bar{r}^{\,5} + 26539\, \bar{r}_h \bar{r}^{\,4} + 28489\, \bar{r}_h^{\,2} \bar{r}^{\,3} 
+ 33379\, \bar{r}_h^{\,3} \bar{r}^{\,2} + 13492\, \bar{r}_h^{\,4} \bar{r} + 4500\, \bar{r}_h^{\,5}\right)
\Bigr\} \alpha^2 +\mathcal{O}(\alpha^3)
\biggr]\!,\label{eq17}
\end{align}
\begin{align}
\bar{h}(\bar{r}) = &\left(1 - \frac{\bar{r}_h}{\bar{r}}\right) \Biggl[ 
1 - \frac{(\bar{r} + 2\,\bar{r}_h)(\bar{r} + 5\,\bar{r}_h)}{18 \bar{M}^{\,2} \bar{r}_h \bar{r}^{\,3}}\, \alpha
- \frac{1}{32400\, M^{\,2} \bar{M}^{\,4} \bar{r}_h^{\,3} \bar{r}^{\,6}}\nonumber\\ 
& \times \Bigl\{\bar{M}^{\,2} \left(21211 \bar{r}^{\,5} + 13231 \bar{r}_h \bar{r}^{\,4} + 11041 \bar{r}_h^{\,2} \bar{r}^{\,3}
- 33019\, \bar{r}_h^{\,3} \bar{r}^{\,2} - 36964\, \bar{r}_h^{\,4} \bar{r} - 40300\, \bar{r}_h^{\,5}\right) e^{-2\phi_0}\nonumber\\ 
& + M^{\,2} (\bar{r} - \bar{r}_h) 
\left(1901 \bar{r}^{\,4} - 4178\, \bar{r}_h \bar{r}^{\,3} - 15747\, \bar{r}_h^{\,2} \bar{r}^{\,2} - 4976\, \bar{r}_h^{\,3} \bar{r} - 1300\, \bar{r}_h^{\,4} \right) \Bigr\} \alpha^2 
+ \mathcal{O}(\alpha^3) \Biggr]\!,\label{eq18}
\end{align}
\begin{align}
\phi(\bar{r}) =\; &\phi_0 
- \frac{(6\,\bar{r}^{\,2} + 3\,\bar{r}_h \bar{r} + 2\,\bar{r}_h^{\,2})(M^{\,2} - \bar{M}^{\,2} e^{-2\phi_0})}{18\, M^{\,2} \bar{M}^{\,2} \bar{r}_h \bar{r}^{\,3}}\, \alpha - \frac{1}{32400\, M^{\,4} \bar{M}^{\,4} \bar{r}_h^{\,3} \bar{r}^{\,6}}\nonumber\\
& \times \Bigl[3 M^{\,4} (740\, \bar{r}^{\,5} - 130\, \bar{r}_h \bar{r}^{\,4} 
- 20\, \bar{r}_h^{\,2} \bar{r}^{\,3} + 1285\, \bar{r}_h^{\,3} \bar{r}^{\,2} + 
6887 \bar{r}_h^{\,4} \bar{r} + 300\, \bar{r}_h^{\,5})\nonumber \\
& - 200\, M^{\,2} \bar{M}^{\,2} e^{-2\phi_0} 
\bigl(33\, \bar{r}^{\,5} + 18\, \bar{r}_h \bar{r}^{\,4} + 
16\, \bar{r}_h^{\,2} \bar{r}^{\,3} + 33\, \bar{r}_h^{\,3} \bar{r}^{\,2} + 
15\, \bar{r}_h^{\,4} \bar{r} + 6\, \bar{r}_h^{\,5} \bigr)\nonumber \\ 
& + \bar{M}^{\,4} e^{-4\,\phi_0} (4380\, \bar{r}^{\,5} + 3990\, \bar{r}_h \bar{r}^{\,4} + 3260\, \bar{r}_h^{\,2} \bar{r}^{\,3} + 2745\, \bar{r}_h^{\,3} \bar{r}^{\,2} + 936\, \bar{r}_h^{\,4} \bar{r} + 300\, \bar{r}_h^{,5}) 
\Bigr] \alpha^2 + \mathcal{O}(\alpha^3).\label{eq19}
\end{align}
Here, $\phi_0$ is the asymptotic value of the scalar field $\phi(\bar{r})$ at 
spatial infinity. Moreover, it is to be noted that the metric functions $f(r)$ 
and $h(r)$ in the coordinate frame defined by the metric tensor $g_{\mu\nu}$ 
can be expressed in terms of the radial coordinate $r$ by utilizing the 
correspondence relation given in Eq.~\eqref{eq9}. Thus by setting 
$\phi_0 = 0$, the resulting expressions for $f(r)$ and $h(r)$ are obtained 
from Eqs.~\eqref{eq17} and \eqref{eq18} through the substitutions of 
$\bar{r} \to r$, $\bar{r}_h \to r_h$, $\bar{M} \to M$, and $M \to \bar{M}$, 
namely:
\begin{align}
f(r) = &\left(1 - \frac{r_h}{r}\right) \Bigg[ 
1 + \frac{23\,r^{\,2} + 11 r_h r + 6\, r_h^{\,2}}{18 M^{\,2} r_h r^{\,3}}\, \alpha - \frac{1}{32400\, \bar{M}^{\,2} M^{\,4} r_h^{\,3} r^{\,6}}\nonumber\\
&\times \Big\{M^{\,2} \left(38731 r^{\,5} + 26371 r_h r^{\,4} + 22721 r_h^{\,2} r^{\,3} - 769\, r_h^{\,3} r^{\,2} - 5572\, r_h^{\,4} r - 9300\, r_h^{\,5} \right) \nonumber \\
& - \bar{M}^{\,2} \left(6979\, r^{\,5} + 26539\, r_h r^{\,4} + 
28489\, r_h^{\,2} r^{\,3} + 33379\, r_h^{\,3} r^{\,2} + 13492\, r_h^{\,4} r + 
4500\, r_h^{\,5} \right)\Big\} \alpha^2 + \mathcal{O}(\alpha^3) 
\Bigg], \label{eq20}
\end{align}
\begin{align}
h(r) = &\left(1 - \frac{r_h}{r} \right) 
\Bigg[1 - \frac{(r + 2r_h)(r + 5r_h)}{18 M^{\,2} r_h r^{\,3}}\, \alpha
- \frac{1}{32400\, \bar{M}^{\,2} M^{\,4} r_h^3 r^{\,6}}\nonumber \\ 
& \times \Big\{ M^{\,2} \left(21211 r^{\,5} + 13231 r_h r^{\,4} + 
11041 r_h^{\,2} r^{\,3} - 33019\, r_h^{\,3} r^{\,2} - 36964 r_h^{\,4} r - 
40300\, r_h^{\,5} \right) \nonumber \\
& + \bar{M}^{\,2} (r - r_h) \left(1901 r^{\,4} - 4178\, r_h r^{\,3} - 
15747 r_h^{\,2} r^{\,2} - 4976\, r_h^{\,3} r - 1300\, r_h^{\,4} \right) 
\Big\} \alpha^2 + \mathcal{O}(\alpha^3)
\Bigg]. \label{eq21}
\end{align}
Thus, Eqs.~\eqref{eq20} and \eqref{eq21} represent the black hole solutions in 
the presence of the GB trace anomaly, accurate up to second order in $\alpha$. 
In the following sections, we will make use of these solutions for further 
investigations. It is to be noted here that in the underlying quantum 
theory, the coupling parameter $\alpha$ is proportional to the GB anomaly 
coefficient of the conformal matter field, which is non-negative for any 
unitary four-dimensional conformal field theory and is therefore 
non-negative in physically consistent anomaly-induced effective actions 
\cite{pos1,pos2,pos3}. For this reason, and following the conventions of 
anomaly-induced effective actions, we restrict our analysis to 
$\alpha \geq 0$ in the present work.
\section{Particle Motion Around Black Holes Modified by Trace Anomaly} 
\label{3}
In this section, we investigate the motion of test particles under different
conditions in the spacetime of BHs influenced by the GB trace anomaly. 
In all numerical computations, we work in geometrized units, 
$G = c = 1$. Following the standard practice in BH accretion and QPO studies, 
we express all radii and frequencies in units of the BH mass $\mathcal{M}$, 
so that these quantities become dimensionless. In practice, we set 
$\mathcal{M} = 1$ for all numerical calculations. Moreover, in the unbarred 
frame, the perturbative condition~\eqref{eq16} takes the form 
$\alpha/M^{2} r_h^{2}$,
where $M$ is the new mass scale appearing in the anomaly-induced action, 
related to the reduced Planck mass by $M = M_{\rm pl}/\sqrt{2}$. In our 
analysis, we consider values of the anomaly parameter $\alpha$ of the order of 
$10^{-5}$, and we identify the horizon radius as $r_h = 2\mathcal{M}$. Thus 
in the chosen unit of $\mathcal{M} = 1$ along with $M_{\rm pl} = 1$, the value 
of $\alpha/M^{2} r_h^{2}$ 
become the order of $\sim10^{-5}$ in the explored parameter range. Since this 
is several orders of magnitude smaller than $1$, the perturbative 
condition~\eqref{eq16} will be safely satisfied throughout the entire parameter 
space used in this work.
\subsection*{A. Equations of Motion}
\begin{figure}[!h]
\includegraphics[scale=0.75]{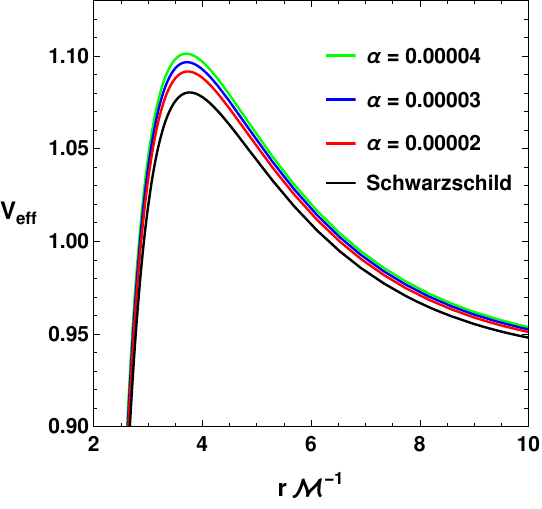}
\vspace{-0.2cm}
\caption{Variation of the effective potential \eqref{eqep} with radial 
coordinate normalized by the BH mass $r\mathcal{M}^{-1}$ for different values 
of $\alpha$. Here we consider $\bar{M} = 0.003$.}
\label{fig1}
\end{figure}

The trajectories of test particles are determined by the following Lagrangian:
\begin{equation}
\mathcal{L}_p = \frac{1}{2}\, m\, g_{\mu\nu} u^{\mu} u^{\nu},
\end{equation}
where $m$ is the mass and $u^\mu = dx^\mu/d\tau$ is the four velocity of a 
test particle. It is important to emphasize that $x^\mu(\tau)$ describes the 
particle’s worldline, which is parametrized by the proper time $\tau$. In a 
spherically symmetric spacetime, the underlying symmetries give rise to two 
Killing vectors: one is associated with time-translation invariance, 
$ \xi^\mu = (1, 0, 0, 0) $, and the other corresponds to rotational symmetry 
about the azimuthal axis, $\eta^\mu = (0, 0, 0, 1) $. These symmetries imply 
the existence of conserved quantities along the particle’s worldline. 
Specifically, they lead to the conservation of the test particle’s total 
energy $\mathcal{E}$ and angular momentum $\mathcal{J}$. These constants of 
motion can be expressed as
\begin{equation}
\mathcal{E} = -\,g_{tt} \dot{t} \quad \text{and} \quad \mathcal{J} = 
g_{\phi \phi} \dot{\phi},
\end{equation}
where the overdot represents differentiation with respect to proper time 
$\tau$. The equation governing the motion of the test particle can be obtained 
by applying the following normalization condition \cite{QPO1}:
\begin{equation}
g_{\mu\nu} u^{\mu} u^{\nu} = \delta, \label{eq25}
\end{equation} 
where $\delta = 0$ for massless particles and $\delta = \pm 1$ for massive 
particles with $\delta = +1$ for the spacelike geodesic of the particle and 
$\delta = -1$ for the timelike geodesic of the particle. Using the above 
condition \eqref{eq25} and considering the motion of the particle in the 
equatorial plane (i.e., $\theta = {\pi}/2$ and $\dot{\theta} = 0$), the 
equations of motion for the test particle around our considered BHs can be 
derived as
\begin{align}\label{eq25a}
\dot{r}^{\,2} & = h(r) \left(\frac{\mathcal{E}^{\,2}}{f(r)} - \frac{\mathcal{J}^2}{r^{\,2}}  - 1 \right),\\[5pt]
\dot{\theta}^2 & = 0,\;\;
\dot{\phi} = \frac{\mathcal{J}}{r^{\,2}},\;\;
\dot{t} = \frac{\mathcal{E}}{f(r)}.
\end{align} 
The radial equation of motion \eqref{eq25a} in terms of coordinate time can be 
expressed in a Newtonian-like form (under the approximation $f(r)\approx h(r)$)
as
\begin{equation}
{\dot{r}}^{\,2} = {\mathcal{E}}^{\,2} - V_\text{eff},
\end{equation}
where $V_\text{eff}$ denotes the effective potential governing the radial 
motion in the equatorial plane, and it is given by
\begin{equation}\label{eqep}
V_\text{eff} = h(r)\left(1 + \frac{{\mathcal{J}}^2}{r^2} \right).
\end{equation}
Fig.~\ref{fig1} illustrates the variation of the effective potential 
$V_\text{eff}$ for a test particle with respect to the radial coordinate 
normalized by the BH mass $r\mathcal{M}^{-1}$ and for different values of the 
GB coupling constant $\alpha$. It should be noted that the values of 
$\alpha$, $\bar{M}$, and $M$ have been chosen to satisfy the 
condition~\eqref{eq16}.
It is evident from Fig.~\ref{fig1} that the peak of the effective potential 
$V_\text{eff}$ increases with increasing values of the GB coupling constant 
$\alpha$, which is higher for all values of $\alpha$ as compared to the 
Schwarzschild BH case. It is to be noted here that in all our numerical
calculations we consider $M_{pl} =1$ and hence $M = 1/\sqrt{2}$.
\subsection*{B. Circular Orbits}
In this subsection, we explore the circular motion of test particles around 
the BHs by imposing the following conditions \cite{QPO1,QPO2}:
\begin{equation}
\dot{r} = 0\, (V_\text{eff} = \mathcal{E}^{\,2})\;\; \text{and}\;\; 
V'_\text{eff} = 0,
\end{equation}
where the prime indicates the derivative with respect to the radial 
coordinate $r$. The condition $V'_\text{eff} = 0$ identifies the stationary 
points of the effective potential, which correspond to circular orbits. The 
requirement that these points be minima ensures the 
stability of the orbits, such that small perturbations will keep the particle 
bound to its circular path. Using these conditions, we derive the expressions 
for the energy and angular momentum of the particle, which are provided in the 
Appendix.
\begin{figure}[!h]
        \centerline{
        \includegraphics[scale = 0.565]{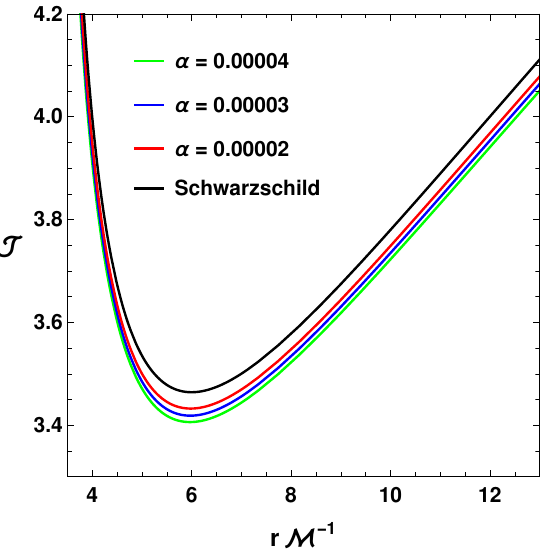}
\hspace{0.3cm}
        \includegraphics[scale = 0.585]{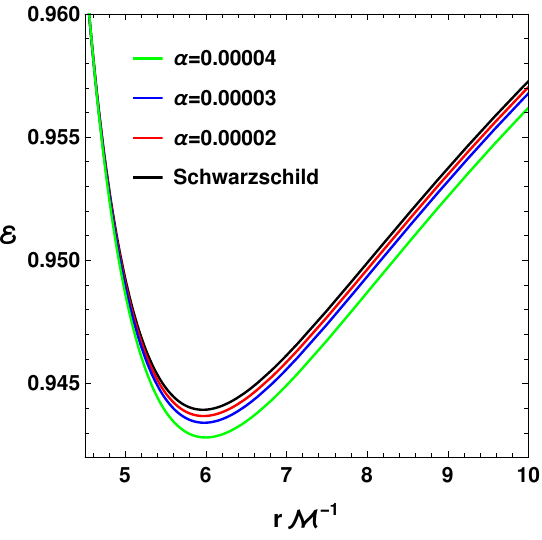}
\hspace{0.3cm}
        \includegraphics[scale = 0.58]{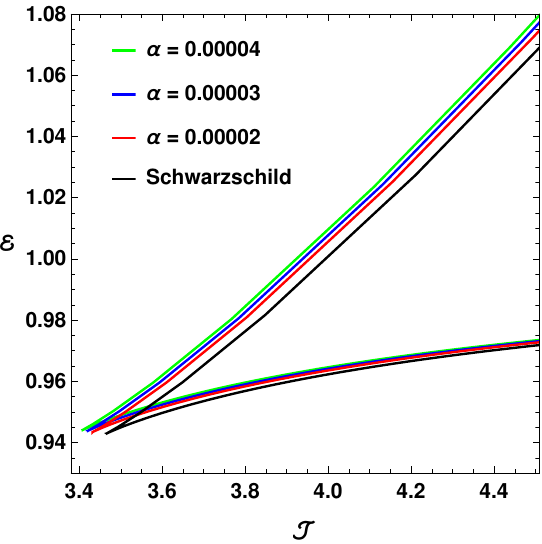}}
        \vspace{-0.2cm}
\caption{Radial profiles of specific angular momentum and energy for circular 
orbits of test particles are shown for different values of $\alpha$, obtained 
by considering $\bar{M} = 0.003$.}
\label{fig2}
\end{figure}
In Fig.~\ref{fig2}, we plot the specific energy $\mathcal{E}$ and 
specific angular momentum $\mathcal{J}$ (not their squared values). Both 
quantities $\mathcal{E}$ and $\mathcal{J}$ are in the unit normalised by the 
BH mass $\mathcal{M}$. This figure depicts the radial dependence of specific 
angular momentum and 
energy of the particle. The left panel shows the variation of the specific 
angular momentum with respect to $r\mathcal{M}^{-1}$. With the increase in the 
value of $\alpha$, the minimum of the angular momentum shifts to lower values 
compared to the Schwarzschild BH. The middle panel shows the variation of the 
energy $\mathcal{E}$, with respect to $r\mathcal{M}^{-1}$. In this case also 
with an increase in the value of the GB trace anomaly coupling constant 
$\alpha$, the minimum of the energy shifts to lower values compared to the 
Schwarzschild BH. The minimum in the energy profile corresponds to the most 
stable circular orbit. The right panel depicts the variation of the specific 
energy $\mathcal{E}$ with specific angular momentum $\mathcal{J}$. As $\alpha$ 
increases, the curves shift, leading to a decrease in the required angular 
momentum for a given energy. Compared to the Schwarzschild case, the presence 
of the trace anomaly allows circular orbits with lower angular momentum for 
the same energy, highlighting the influence of the GB trace anomaly on 
particle dynamics near BHs. 
\subsection*{C. Innermost Stable Circular Orbits}
The stable circular orbits for a test particle around a BH exist at a specific 
radius, denoted by $r = r_\text{min}$, which corresponds to the point where the 
particle possesses the least possible energy and angular momentum required 
to maintain a circular trajectory. This particular radius represents an 
energetically favorable position for the particle's motion. When considering 
the innermost stable circular orbit (ISCO), an additional criterion must be 
met to ensure the stability of the orbit against small perturbations. 
Specifically, the effective potential $V_\text{{eff}}$ governing the radial 
motion of the particle must satisfy the conditions that both the first and 
second derivatives of its must vanish, i.e., $V'_\text{eff} = 0$ and 
$V''_\text{eff} = 0$. These two conditions together determine the precise 
location of the ISCO, ensuring that the orbit is not only circular but also 
marginally stable. Since the expression for the effective potential in our 
chosen BH system involves complicated mathematical terms, we used numerical 
methods to determine the ISCO radius. The variation of the ISCO radius with 
respect to the GB coupling constant $\alpha$ is then plotted for different 
values of the parameter $\bar{M}$.
\begin{figure}[!h]
\includegraphics[scale=0.7]{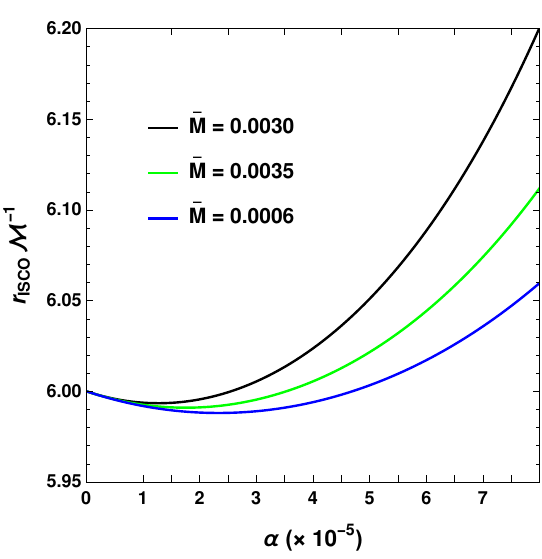}
\vspace{-0.2cm}
\caption{Variation of the ISCO radius normalized by the BH mass 
$\mathcal{M}$ with respect to GB coupling constant $\alpha$ for different 
values of $\bar{M}$.}
\label{fig3}
\end{figure}
Fig.~\ref{fig3} illustrates the variation of the ISCO radius normalized by the 
BH mass with respect to $\alpha$. It is observed that as $\alpha$ increases, 
the ISCO radius also increases for all chosen values of $\bar{M}$. This trend 
suggests that the effect of the GB trace anomaly is to push the stable 
circular orbits outward, causing particles to orbit at relatively larger radii 
as $\alpha$ grows. Additionally, for higher values of $\bar{M}$, the ISCO 
radius is consistently larger, indicating that both $\alpha$ and $\bar{M}$ 
contribute to shifting the ISCO outward. This behavior reflects the impact of 
higher-curvature corrections on the geodesic structure of the BH spacetime.
\section{Fundamental Frequencies} \label{4}
In this section, we derive the analytical expressions for the fundamental 
frequencies associated with the motion of test particles around BHs influenced 
by the GB trace anomaly. These frequencies are essential for characterizing 
small perturbative oscillations about circular geodesics in the BH spacetime. 
Such oscillations provide a theoretical basis for interpreting QPOs observed 
in the X-ray flux of accreting compact objects. The study of these fundamental 
frequencies, specifically the orbital (Keplerian), radial epicyclic, and 
vertical epicyclic modes, serves as a powerful diagnostic tool to probe the 
nature of strong-field gravity near the event horizon. Within the framework of 
GB trace anomaly gravity, which introduces quantum corrections via a trace 
anomaly term in the effective gravitational action, deviations from  GR are 
naturally incorporated. Consequently, analyzing these oscillation modes in 
the modified geometry offers a pathway to constrain the anomaly coupling 
parameter.
\subsection*{A. Keplerian Frequencies}
\begin{figure}[!h]
\includegraphics[scale=0.7]{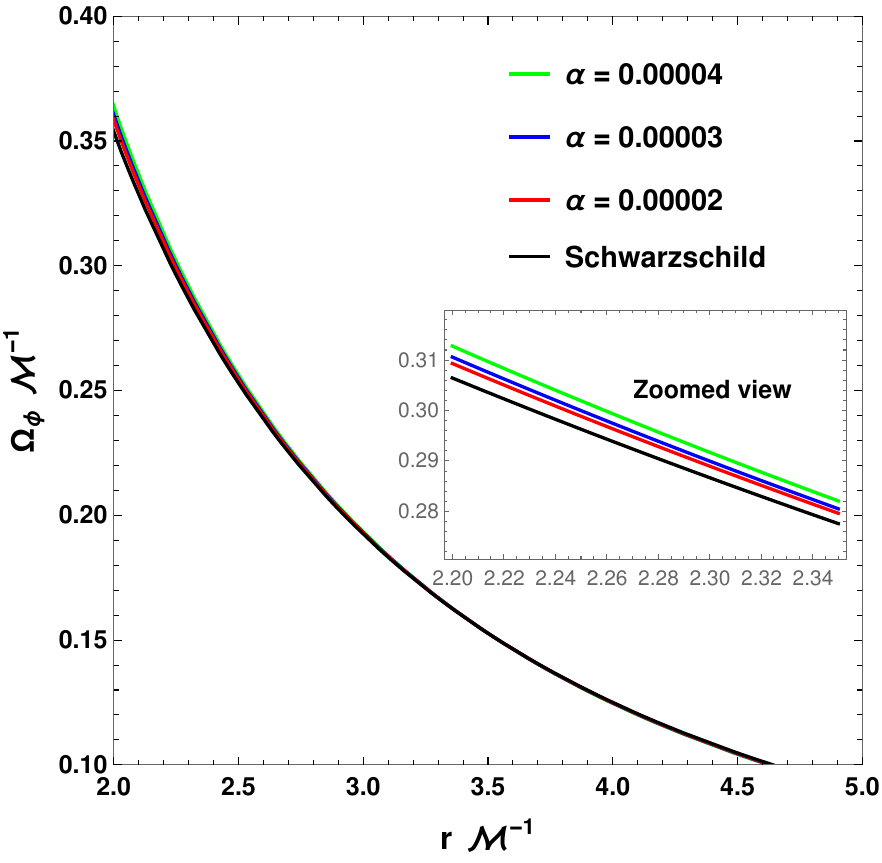}
\vspace{-0.2cm}
\caption{Variation of $\Omega_{\phi}\mathcal{M}^{-1}$ with respect to 
$r\mathcal{M}^{-1}$ for different values of $\alpha$ obatined by considering 
$\bar{M} = 0.003$.} \label{fig5}
\label{fig4}
\end{figure}

The Keplerian frequency refers to the rate at which a test particle moves 
around a BH, as measured by an observer located far from the gravitational 
source. It is commonly represented by $\Omega_{\phi}$, and is defined as the 
ratio of the change in the azimuthal coordinate to the change in coordinate 
time, i.e., $\Omega_{\phi} = d\phi/dt$. Based on this definition, a 
general formula for the orbital or Keplerian frequency can be obtained in the 
context of a static and spherically symmetric spacetime geometry as 
follows~\cite{QPO3}:
\begin{equation}
\Omega_{\phi} = \sqrt{\frac{-\partial_{r} g_{tt}}{\partial_{r} g_{\phi \phi}}} = \sqrt{\frac{f'(r)}{2r}}.
\end{equation} 
The Keplerian frequency for the Schwarzschild BH has the following form 
\cite{QPO1}:
\begin{equation}
\Omega_{\phi} = \sqrt{\frac{\mathcal{M}}{r^3}}. \label{eq33}
\end{equation}
We derive the Keplerian frequency $\Omega_{\phi}$ for the BHs endowed with 
GB trace anomaly, as presented in Eq.~\eqref{omegaphi} in the appendix. Taking 
the limit $\alpha \to 0$ in Eq.~\eqref{omegaphi}, we recover the result given 
in Eq.~\eqref{eq33}. Fig.~\ref{fig4} illustrates the variation of the Keplerian 
frequency, normalized by the BH mass $\mathcal{M}$, as a function of the 
normalized radial coordinate $r\mathcal{M}^{-1}$.
It is evident from Fig.~\ref{fig4} that the magnitude of the Keplerian 
frequency $\Omega_{\phi}\mathcal{M}^{-1}$ decreases monotonically with 
increasing $r \mathcal{M}^{-1}$, as expected for circular orbits around a BH. 
When the GB trace anomaly correction is included, the resulting frequency 
profiles remain very close to the Schwarzschild curve for 
$r \mathcal{M}^{-1} > r_h \mathcal{M}^{-1}$ ($r_h \mathcal{M}^{-1} = 2$ in our case). This behaviour reflects the 
fact that the anomaly induced modification to the metric is a small 
perturbation that falls off rapidly with radius.

To express the angular frequency in terms of the physical frequency measured 
in Hertz (Hz), we use the relation as given by
\begin{equation}\label{con}
\nu_{\phi} = \frac{1}{2\pi} \frac{c^3}{G \mathcal{M}}\, \Omega_{\phi}\; \text{[Hz]}.
\end{equation}
To convert the frequencies from the geometrized units to Hertz in the 
International System of Units (SI), we use the speed of light in vacuum, 
$c = 3 \times 10^8~\mathrm{m\, s^{-1}}$, and the gravitational constant, 
$G = 6.67 \times 10^{-11}~\mathrm{m^3\,kg^{-1}\,s^{-2}}$.
\subsection*{B. Harmonic Oscillations}
In this subsection, we investigate the fundamental frequencies associated with 
the oscillatory motion of test particles around the BHs influenced by the GB 
trace anomaly. These fundamental frequencies, corresponding to radial and 
vertical oscillations, are determined by introducing small perturbations to 
the circular motion of the particles, especially by considering deviations of 
the form: $r \rightarrow r_0 + \delta r$ for radial perturbations and 
$\theta \rightarrow \theta_0 + \delta \theta$ for vertical perturbations. To 
facilitate this analysis, the effective potential is expanded in terms of 
$r$ and $\theta$ around the equilibrium circular orbit as follows:
\begin{align}
V_{\mathrm{eff}}(r, \theta) =\ & V_{\mathrm{eff}}(r_0, \theta_0) + \delta r\, \left. \frac{\partial V_{\mathrm{eff}}(r, \theta)}{\partial r} \right|_{r_0, \theta_0} + \delta \theta\, \left. \frac{\partial V_{\mathrm{eff}}(r, \theta)}{\partial \theta} \right|_{r_0, \theta_0} + \frac{1}{2} \delta r^2\, \left. \frac{\partial^2 V_{\mathrm{eff}}(r, \theta)}{\partial r^2} \right|_{r_0, \theta_0} \nonumber \\[5pt]
& + \frac{1}{2} \delta \theta^2\, \left. \frac{\partial^2 V_{\mathrm{eff}}(r, \theta)}{\partial \theta^2} \right|_{r_0, \theta_0} + \delta r\, \delta \theta\, \left. \frac{\partial^2 V_{\mathrm{eff}}(r, \theta)}{\partial r \partial \theta} \right|_{r_0, \theta_0} + \mathcal{O}(\delta r^3, \delta \theta^3) \,. \label{eq35}
\end{align}
By enforcing the conditions for circular motion and its stability, the 
expansion of the effective potential simplifies such that only the 
second-order derivative terms contribute to the dynamics. As a result, the 
radial and vertical perturbations reduce to equations of harmonic oscillators 
in the equatorial plane. Consequently, these perturbations give rise to 
characteristic frequencies, which correspond to observable quantities for a 
distant observer. The forms of the reduced harmonic oscillator equations are 
as follows \cite{QPO1}:
\begin{align}
\frac{d^2 \delta r}{dt^2} + \Omega_r^2 \delta r &= 0, \quad
\frac{d^2 \delta \theta}{dt^2} + \Omega_\theta^2 \delta \theta = 0, \tag{24}
\end{align}
where $\Omega_r$ and $\Omega_{\theta}$ respectively, are the radial and 
vertical oscillation frequencies and can be express as follows: 
\begin{align}
\Omega_r^2 & = -\,\frac{1}{2 g_{rr} \, \dot{t}^2}\,  {\partial_{r}}^2 V_\text{eff}(r, \theta)|_{\theta = \pi/2},\\[5pt]
\Omega_{\theta}^2 & = -\,\frac{1}{2g_{\theta \theta}{\dot{t}}^2}\, {\partial_{\theta}}^2 V_\text{eff} \left(r, \theta\right) |_{\theta = \pi/2}.
\end{align}
For the BHs under consideration, the expressions for the radial and vertical 
frequencies are given by Eqs.~\eqref{Omega} and \eqref{omegaphi} respectively 
in the appendix. These frequencies can be expressed in terms of physical 
frequencies measured in Hertz using the relation \eqref{con} by generalising 
$\Omega_{\phi}$ and $\nu_\phi$ as $\Omega_i$ and $\nu_i$, where $i$ stands 
for both $r$ and $\theta$ as well as $\phi$.
\section{QPO models and QPO orbits in Black Holes}\label{5}
\subsection*{A. QPO models}
This subsection investigates the possible frequencies of twin-peak QPOs in the 
spacetime of BHs influenced by the GB trace anomaly, providing a comparison 
with the corresponding results for standard Schwarzschild BHs across various 
QPO models. We present the forms of the upper $\left(\nu_{U}\right)$ 
and lower $\left(\nu_{L}\right)$ QPO frequencies in terms of $\nu_r$, 
$\nu_\theta$ and $\nu_\phi$, formulated according to each specific QPO 
model under consideration \cite{QPO4}. For instance, we adopt the following 
commonly used models.
\subsubsection{Parametric Resonance (PR) Model}
The frequent detection of a 3:2 ratio in twin-peak high-frequency QPOs from 
BH and neutron star systems has motivated the proposal that these oscillations 
may originate from nonlinear resonances between different modes of motion in 
the accretion disk \cite{QPO4,QPO5,QPO6}. In its simplest formulation, small 
perturbations in the radial and vertical directions around circular, 
equatorial geodesics are treated as independent harmonic oscillations, 
characterized by the radial ($\nu_r$) and vertical ($\nu_\theta$) epicyclic 
frequencies, respectively. The PR model posits that radial oscillations, 
typically stronger than their vertical counterparts due to 
$\delta r > \delta \theta$ in thin accretion disks, can parametrically excite 
vertical oscillations when a specific resonance condition is met, given by 
$\nu_r / \nu_\theta = 2/n$, where $n$ is a positive integer. In the case of 
rotating BHs, where $\nu_\theta > \nu_r$ generally holds, the resonance is 
strongest for $n = 3$, naturally leading to the frequently observed 3:2 
frequency ratio. Within this framework, the upper and lower QPO frequencies 
are identified as $\nu_U = \nu_\theta$ and $\nu_L = \nu_r$, respectively.
\subsubsection{Relativistic Precession (RP) model}
In the RP model, QPOs in X-ray binaries are interpreted as a natural 
consequence of matter orbiting in the curved spacetime around compact 
objects. Here, plasma blobs in the accretion disk follow slightly eccentric 
and inclined geodesic orbits near the BHs, giving rise to distinct 
frequencies. According to this model, the upper kHz QPO corresponds to the 
Keplerian frequency, given by $\nu_U = \nu_\phi$, while the lower kHz QPO is 
associated with the periastron precession frequency, which is given by 
$\nu_L = \nu_\phi - \nu_r$. Additionally, low-frequency QPO frequencies are 
linked to nodal precession, given by $\nu_\phi - \nu_\theta$, related to 
vertical oscillations and frame dragging. This nodal frequency vanishes in 
the Schwarzschild case as in the Schwarzschild case $\nu_\phi = \nu_\theta$. 
These frequency identifications are strongly supported by general relativistic 
corrections to orbital motion, including effects of frame dragging and 
spacetime curvature, which naturally give rise to the observed precessional 
phenomena. Furthermore, the detection of harmonic structures in both neutron 
star and BH systems reinforces this interpretation, particularly in cases 
where even harmonics of the nodal precession frequency dominate the observed 
power spectrum \citep{QPO9,QPO10,QPO11}. The RP model provides a relativistic, 
geometrical explanation for QPOs, without invoking resonance mechanisms or 
magnetic fields.
\subsubsection{Warped Disk (WD) model}
The WD model explains high-frequency QPOs by considering a non-standard, 
warped geometry of the accretion disk \cite{QPO12,QPO13}. In this framework, 
nonlinear resonances occur between the warped structure of the disk and its 
intrinsic oscillation modes. These resonances involve interactions in both 
horizontal and vertical directions. The horizontal resonances can excite both 
g-mode and p-mode oscillations, whereas vertical resonances typically excite 
only g-modes. The emergence of these resonances is attributed to the 
non-monotonic variation of the radial epicyclic frequency with respect to the 
radial coordinate $r$. According to this model, the observed high-frequency 
QPOs are associated with the relations $\nu_U = 2\nu_\phi - \nu_r$ and 
$\nu_L = 2(\nu_\phi - \nu_r)$.
%
\subsubsection{Tidal Disruption (TD) Model}
The TD model interprets QPOs as a consequence of the tidal breakup of 
inhomogeneities, such as dense clumps of matter or external bodies, in the 
strong gravitational field of a BH. According to this model, the observed QPO 
frequencies are linked to the fundamental orbital frequencies of the disrupted 
material. Specifically, the upper frequency is identified as 
$\nu_U = \nu_\phi + \nu_r$, while the lower frequency corresponds to 
$\nu_L = \nu_\phi$. The TD model has been motivated by numerical simulations of 
tidal disruption events, where compact objects such as asteroids or stars are 
torn apart by the BH tidal forces. These simulations reveal the formation of 
a transient, bright, ring-like structure featuring an orbiting radiating core. 
This dynamically evolving configuration can naturally give rise to QPOs 
through the superposition of the orbital and epicyclic motions of the 
disrupted material. Thus, the TD model offers a physically motivated, 
relativistic mechanism for QPO generation based on the extreme tidal 
interactions near BHs \cite{QPO4,QPO14}.
\begin{figure}[!h]
    \centerline{
     \includegraphics[scale=0.613]{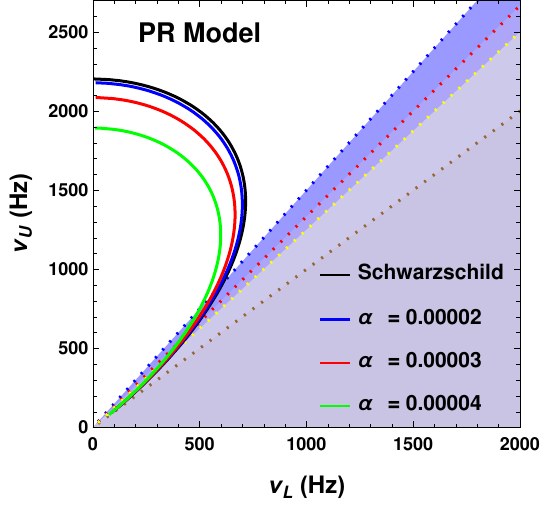}\hspace{0.3cm}
     \includegraphics[scale=0.593]{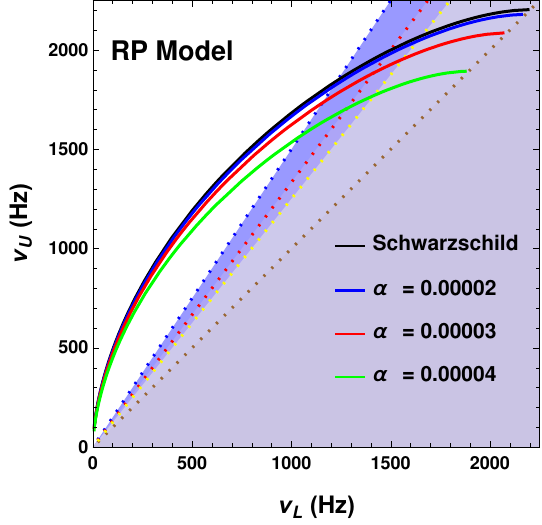}\hspace{0.3cm}
     \includegraphics[scale=0.593]{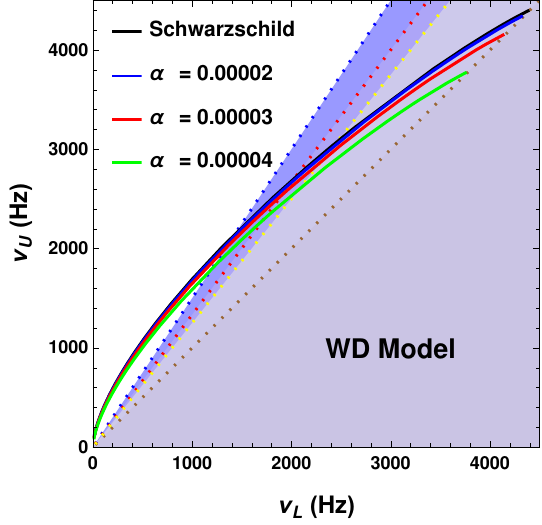}
     }\vspace{0.4cm}
     \centerline{
     \includegraphics[scale=0.593]{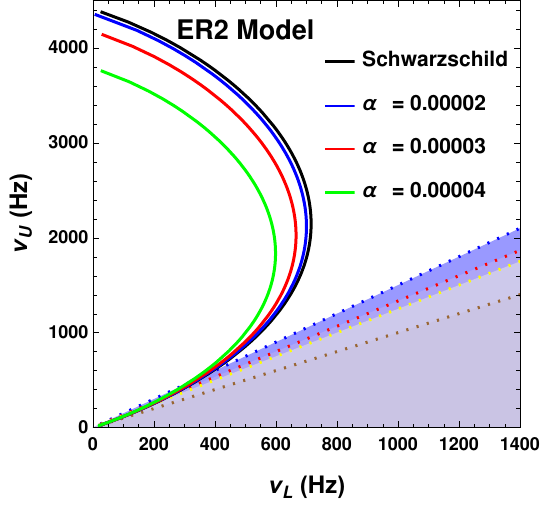}\hspace{0.3cm}
     \includegraphics[scale=0.593]{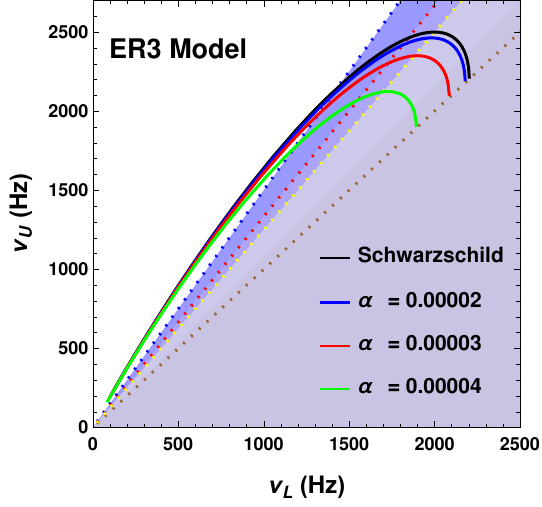}\hspace{0.3cm}
     \includegraphics[scale=0.593]{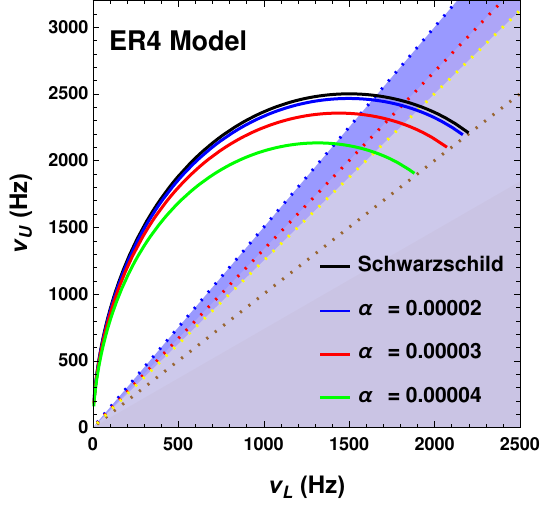}}\vspace{0.4cm}
     \centerline{\includegraphics[scale=0.593]{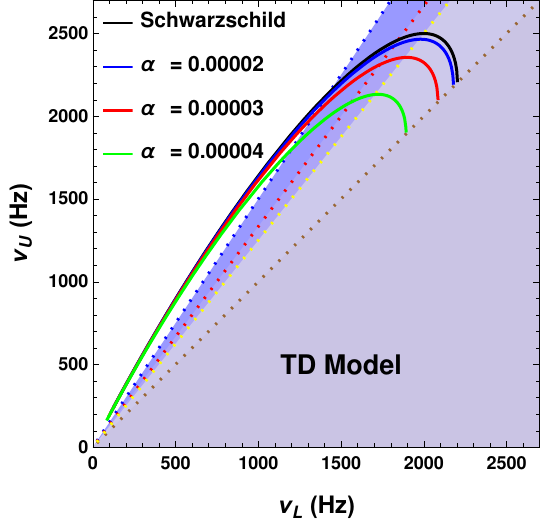}}
    \vspace{-0.2cm}
    \caption{Correlations between the upper ($\nu_U$) and lower ($\nu_L$)
frequencies of twin-peak QPOs for different values of the coupling constant
$\alpha$ as predicted by different QPO models mentioned in the text. In this
analysis, we consider $\bar{M} = 0.003$ as a fixed parameter value.}
\label{fig5}
\end{figure}
\subsubsection{ER2, ER3, and ER4 Models}
The epicyclic resonance (ER) model offers a relativistic framework for 
interpreting high-frequency QPOs as the result of nonlinear resonances between 
the fundamental oscillation modes of accreting matter in the vicinity of 
compact objects. In this scenario, QPOs arise due to resonances between the 
radial and vertical epicyclic frequencies of particles orbiting along slightly 
perturbed geodesic trajectories in a relativistic accretion disk. In this 
work, we specifically focus on three representative cases of the ER model, 
viz., ER2, ER3, and ER4, each characterized by distinct combinations of the 
orbital and epicyclic frequencies. The QPO frequencies for these models are 
defined as follows \cite{QPO1}:
\begin{itemize}
\item For ER2, the upper and lower QPO frequencies are given by 
$\nu_U = 2\nu_\theta - \nu_r$ and $\nu_L = \nu_r$.
\item For ER3, they are expressed as $\nu_U = \nu_\theta + \nu_r$ and 
$\nu_L = \nu_\theta$.
\item For ER4, the relations take the form: 
$\nu_U = \nu_\theta + \nu_r$ and $\nu_L = \nu_\theta - \nu_r$.
\end{itemize}

Fig.~\ref{fig5} shows the relationship between the computed values of $\nu_U$ 
and $\nu_L$ for the considered BH spacetime, evaluated using the PR, RP, WD, 
TD, and ER2–ER4 models for different values of the GB coupling constant 
$\alpha$. The diagram includes characteristic reference lines corresponding to 
the frequency ratios 3:2, 4:3, 5:4, and 1:1 between the upper and lower 
frequencies. Specifically, the 3:2 ratio is represented by a dotted blue line, 
4:3 by a dotted red line, 5:4 by a dotted yellow line, and 1:1 by a dotted 
brown line. In addition, the plot shows shaded regions between these reference 
lines, visually highlighting distinct frequency ratio bands for the twin-peak 
QPOs. The line corresponding to the 1:1 frequency ratio is commonly referred 
to as the QPO graveyard, as any QPO located on this line indicates the merging 
of the upper and lower frequencies into a single oscillation, effectively 
erasing the characteristic twin-peak structure. Fig.~\ref{fig5} clearly 
illustrates that as the GB coupling constant $\alpha$ increases, the predicted 
relationship between the upper and lower QPO frequencies progressively departs 
from the Schwarzschild reference curve, emphasizing the impact of the GB trace 
anomaly on the QPO characteristics.
\subsection*{B. QPO Orbits} \label{subB}
\begin{figure}[!h]
    \centerline{
     \includegraphics[scale=0.59]{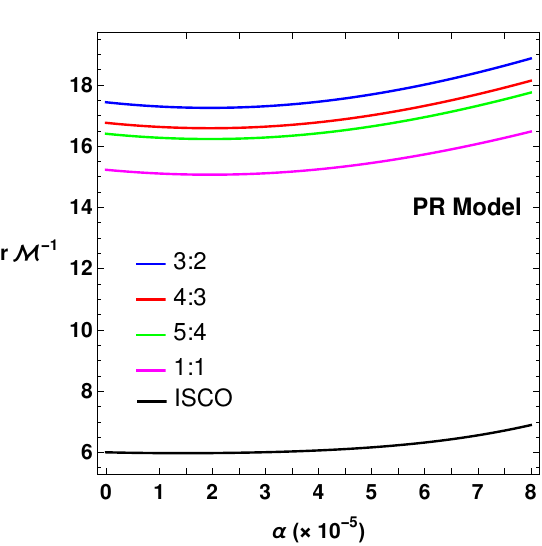}\hspace{0.3cm}
     \includegraphics[scale=0.595]{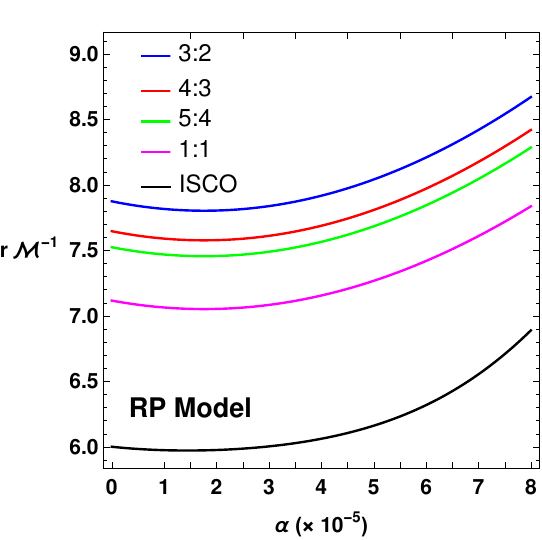}\hspace{0.3cm}
     \includegraphics[scale=0.59]{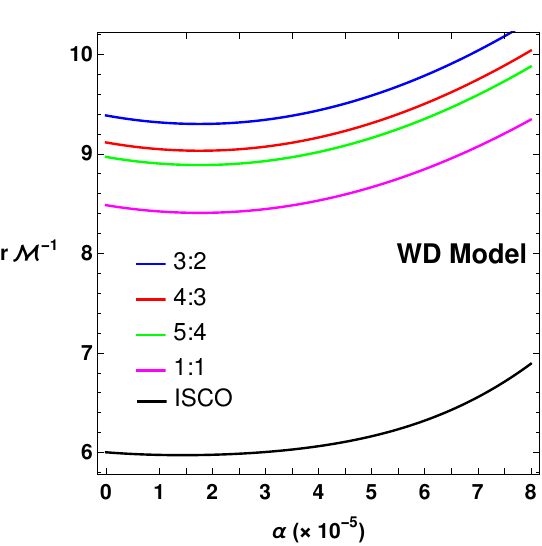}
     }\vspace{0.3cm}
     \centerline{
     \includegraphics[scale=0.60]{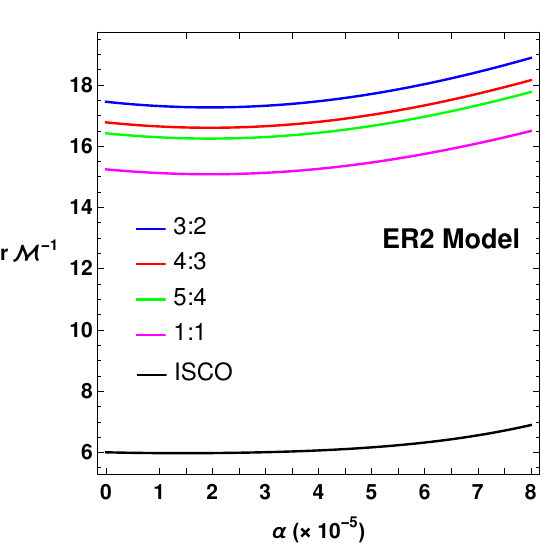}\hspace{0.3cm}
     \includegraphics[scale=0.60]{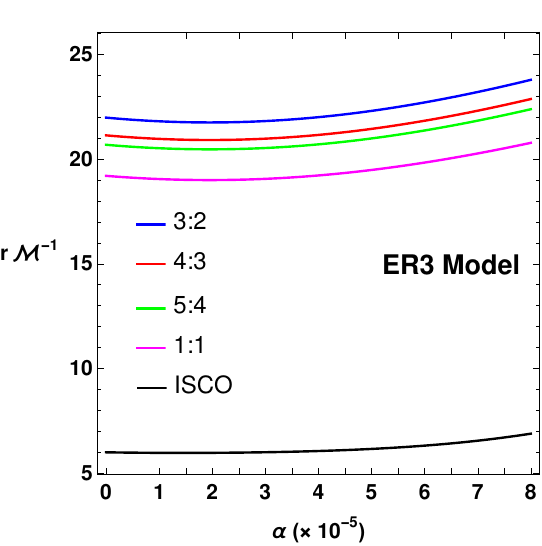}\hspace{0.3cm}
     \includegraphics[scale=0.60]{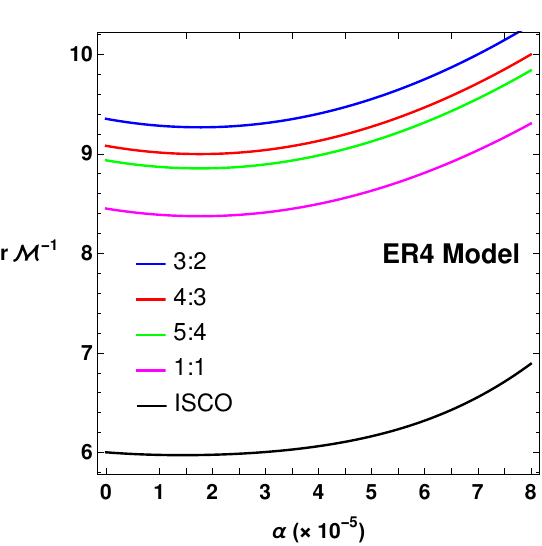}}\vspace{0.3cm}
     \centerline{\includegraphics[scale=0.63]{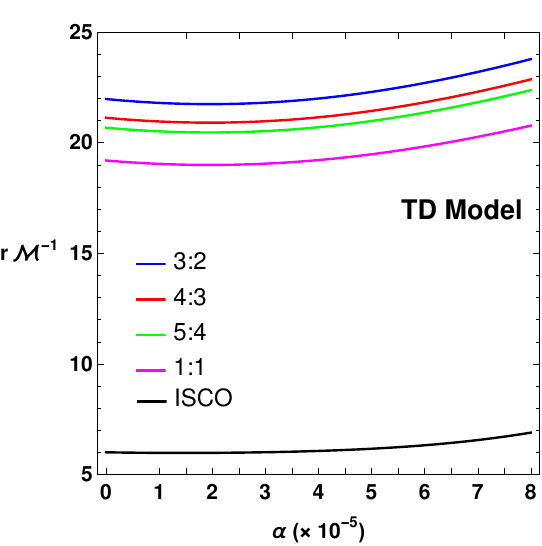}}
     \vspace{-0.2cm}
    \caption{Dependence of the QPO orbital radius on the GB coupling 
constant $\alpha$ for PR, RP, WD, ER2, ER3, ER4 and TD models.}
\label{fig6}
\end{figure}

In this subsection, we explore the influence of the GB coupling constant 
$\alpha$ on the orbital radii associated with QPOs, particularly those 
characterized by frequency ratios 1:1, 3:2, 4:3, and 5:4. These characteristic 
radii can be determined by satisfying the resonance condition \cite{QPO1}:
\begin{equation}
a \nu_{U} \left(\mathcal{M}, r, \alpha \right) = b \nu_{L} \left(\mathcal{M}, r, \alpha \right), \label{eq41}
\end{equation}
where $a$ and $b$ are integers that define the specific resonance ratio, and 
$\nu_U$, $\nu_L$ represent the upper and lower QPO frequencies respectively, 
as mentioned already. This condition is solved numerically for the radial 
coordinate $r\mathcal{M}^{-1}$ across various values of the GB parameter 
$\alpha$, within the frameworks of the QPO models: PR, RP, WD, TD and 
ER2--ER4. The resulting solutions for $r\mathcal{M}^{-1}$ are presented in 
Fig.~\ref{fig6}, which illustrates how the resonance locations evolve as a 
function of the coupling constant $\alpha$.
Fig.~\ref{fig6} displays the dependence of the QPO-generating orbital radii 
expressed in the unit of $\mathcal{M}$ on the GB coupling constant $\alpha$ 
for all seven QPO models considered: PR, RP, WD, ER2, ER3, ER4, and TD. Each 
panel corresponds to a specific model and presents the orbital locations 
associated with the resonant frequency ratios 1:1, 3:2, 4:3, and 5:4, along 
with the ISCO. The plots reveal that the ER3 and TD models yield QPOs at 
significantly larger orbital radii compared to the other models, indicating 
the most extended resonance region among those models. This is followed by the 
ER2 and PR models, which also predict large resonance radii. In contrast, the 
RP model consistently produces the smallest QPO orbital radii. Furthermore, a 
consistent trend is observed across all models that the QPOs generating radii 
increase monotonically with the GB coupling parameter $\alpha$, suggesting that 
the presence of the GB correction modifies the underlying spacetime geometry.
\section{Monte Carlo Markov Chain Analysis} \label{6}
In this section, we employ the Markov Chain Monte Carlo (MCMC) technique to 
place constraints on the parameters of our chosen BH system, using 
observational data from six prominent sources that span three distinct mass 
scales. These include four stellar-mass BH: GRO J1655–40, XTE J1550–564, 
GRS 1915+105, and H 1743–322, one intermediate-mass BH, M82 X-1, and the 
supermassive BH Sgr A*. A concise summary of these BH sources along with their 
corresponding observational characteristics is provided in Table~\ref{table1} 
\cite{mc1,mc2,mc3,mc4,mc5,mc6,mc7,mc8,mc9,mc10,mc11}. In this work, we use the 
RP model as a representative case to constrain the BH parameters by performing 
an MCMC analysis on these six BHs. 
\begin{table}[!h]
\caption{Mass and QPO frequencies of selected BH sources.}
\vspace{5pt}
\centering
\begin{tabular}{|c|c|c|c|}
\hline
\textbf{BH} & \textbf{Mass (in $M_\odot$)} & \textbf{Upper Frequency (Hz)} & \textbf{Lower Frequency (Hz)} \\
\hline
GRO J1655--40   & $5.4 \pm 0.3$  & $441 \pm 2$  & $298 \pm 4$ \\ \hline
XTE J1550--564  & $9.1 \pm 0.61$ & $276 \pm 3$  & $184 \pm 5$ \\ \hline
GRS 1915+105    & $12.4^{+2.0}_{-1.8}$  & $168 \pm 3$  & $113 \pm 5$ \\
\hline
H 1743+322      & $8.0$ -- $14.07$ & $242 \pm 3$ & $166 \pm 5$ \\ \hline
M82 X-1         & $415 \pm 63$  & $5.07 \pm 0.06$ & $3.32 \pm 0.06$\\\hline
Sgr A*          & $(3.5$ -- $4.9) \times 10^6$  & $(1.445 \pm 0.16) \times 10^{-3}$ & $(0.886 \pm 0.04) \times 10^{-3}$ \\
\hline
\end{tabular}  \label{table1}
\end{table}

In order to get meaningful constraints on the model parameters, we use a 
Bayesian statistical framework. This approach allows us to incorporate prior 
knowledge and observational data in a consistent manner using the MCMC 
technique. The Bayesian posterior distribution is given by \cite{mc12,mc13} 
\begin{equation}
P(\boldsymbol{\theta} | D, \mathcal{T}_m) = \frac{P(D | \boldsymbol{\theta}, \mathcal{T}_m)\, \pi(\boldsymbol{\theta} | \mathcal{T}_m)}{P(D | \mathcal{T}_m)},
\label{eq42}
\end{equation}
where $\boldsymbol{\theta}$ is the set of parameters in the model, i.e.,
$\boldsymbol{\theta} = (\mathcal{M}, \bar{M}, \alpha, r\mathcal{M}^{-1})$ for 
our case,  $D$ is the observed data, $\mathcal{T}_m$ denotes the theoretical 
model, $P(D | \boldsymbol{\theta}, \mathcal{T}_m)$ is the likelihood function, 
which measures how well the model with parameters $\boldsymbol{\theta}$ 
predicts the observed data $D$, $\pi(\boldsymbol{\theta} | \mathcal{T}_m)$ 
represents the prior distribution and $P(D | \mathcal{T}_m)$ is the marginal 
likelihood, which is a normalization factor ensuring the total probability 
over all $\boldsymbol{\theta}$ sums to $1$. A Gaussian prior can be defined as
\begin{equation}
\pi(\theta_i) \exp\left[\frac{1}{2} \left(\frac{\theta_i - \theta_{0,i}}{\sigma_i}\right)^2\right]\!,\; \theta_{low,i} < \theta_i < \theta_{high,i}, \label{eq43}
\end{equation}
where $\theta_i = (\mathcal{M}, \bar{M}, \alpha, r_i\mathcal{M}^{-1})$, $\theta_{0,i}$ is the mean and 
$\sigma_i$ are the deviations of the corresponding parameters. In this work, 
we consider the priors to have Gaussian distributions.

The log-likelihood function is defined as the natural logarithm of the 
likelihood function. In the present case, it can be expressed as
\begin{equation}
\log P = \log P_U + \log P_L.
\end{equation}
Here, $\log P_U$ is the log-likelihood of upper frequency and 
$\log P_L$ is the log-likelihood of lower frequency. These two 
likelihoods can be expressed as follows:
\begin{align}
\log P_U & = -\,\frac{1}{2} \sum_i \frac{(\nu_{\phi,obs}^{i} - \nu_{\phi,th}^{i})^2}{(\sigma_{\phi,obs}^i)^2}, \label{eq45}\\[5pt]
\log P_L & = -\,\frac{1}{2} \sum_i \frac{(\nu_{L,obs}^{i} - \nu_{L,th}^{i})^2}{(\sigma_{L,obs}^i)^2}. \label{eq46}
\end{align}
The quantities $\nu_{\phi,\text{obs}}$ and $\nu_{L,\text{obs}}$ represent the 
observed values of the orbital (Keplerian) frequency and the periastron 
precession (lower) frequency, respectively, for the astrophysical source under 
consideration. The lower frequency $\nu_L$ is calculated as the difference 
between the Keplerian frequency $\nu_\phi$ and the radial epicyclic frequency 
$\nu_r$. Correspondingly, $\nu_{\phi,\text{th}}$ and $\nu_{L,\text{th}}$ 
denote the theoretical predictions for these frequencies based on the chosen 
model. 

We then carry out an MCMC analysis to estimate the parameters of our chosen 
BHs. For this, we use known values from earlier QPO studies as a guide. For 
each parameter, we generate around $10^5$ samples from its respective Gaussian 
prior. This approach enables a thorough exploration of the physically relevant 
parameter space and helps in determining the most suitable parameter values. 
\begin{figure}[!h]
    \centerline{
     \includegraphics[scale=0.285]{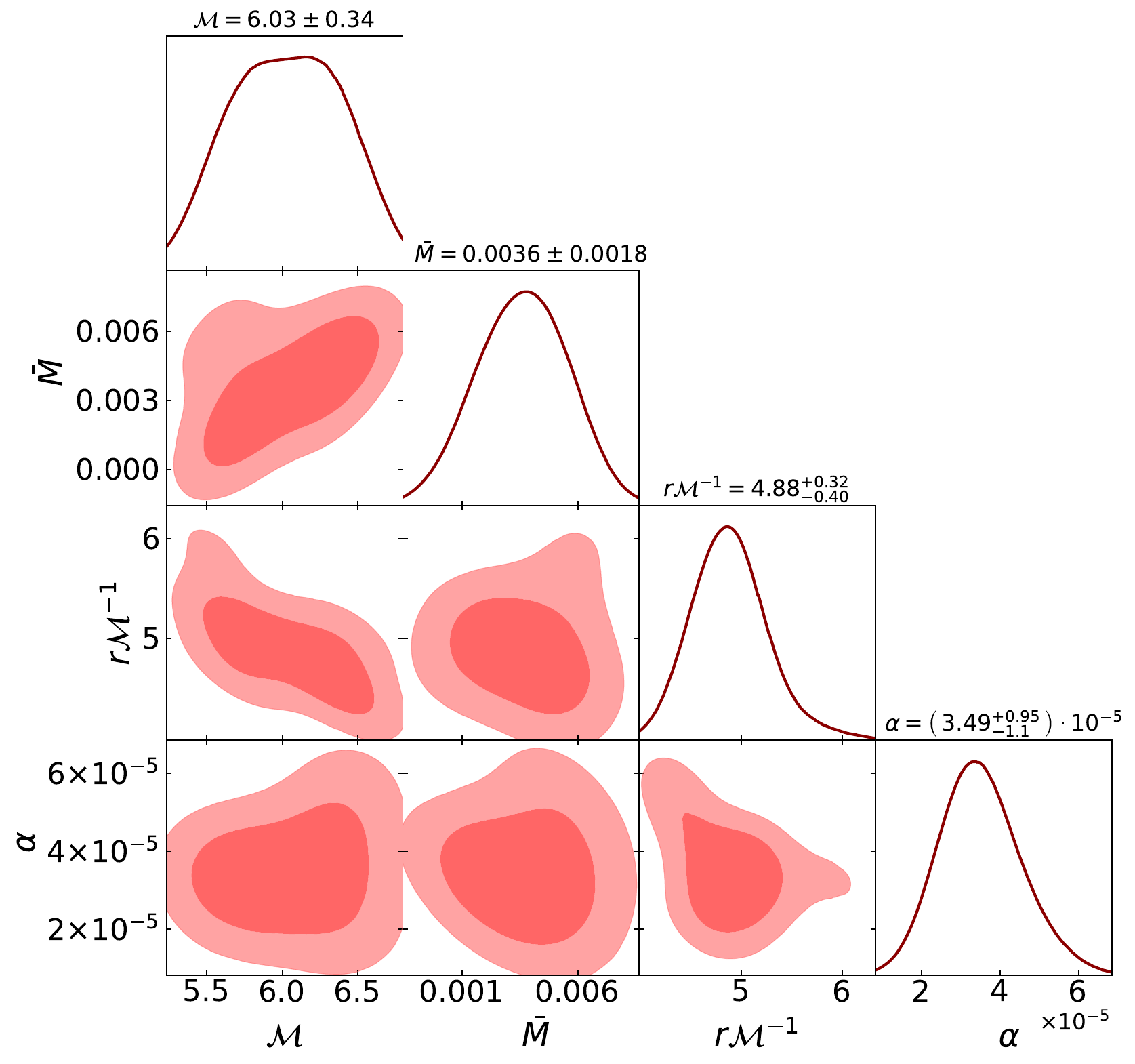}\hspace{0.3cm}
     \includegraphics[scale=0.285]{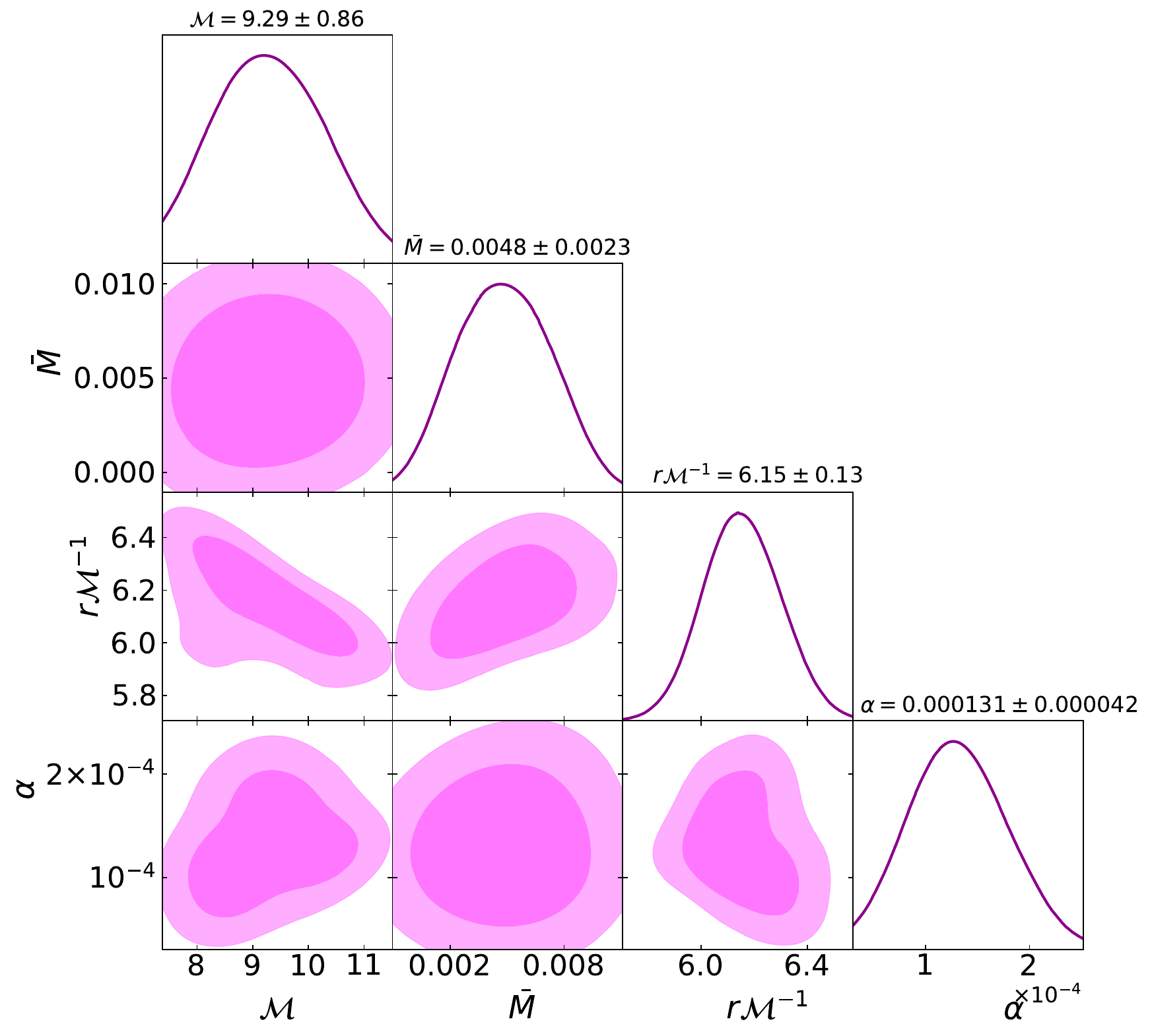}}
     \vspace{-0.2cm}
     \caption{MCMC outputs, showing the constraints on the parameters 
$\mathcal{M}$, $\bar{M}$, $r\mathcal{M}^{-1}$ and $\alpha$, using data from 
GRO J1655–40 (left) and XTE J1550–564 (right) BHs.}
\label{fig7}
\end{figure}
\begin{figure}[!h]
     \centerline{
     \includegraphics[scale=0.285]{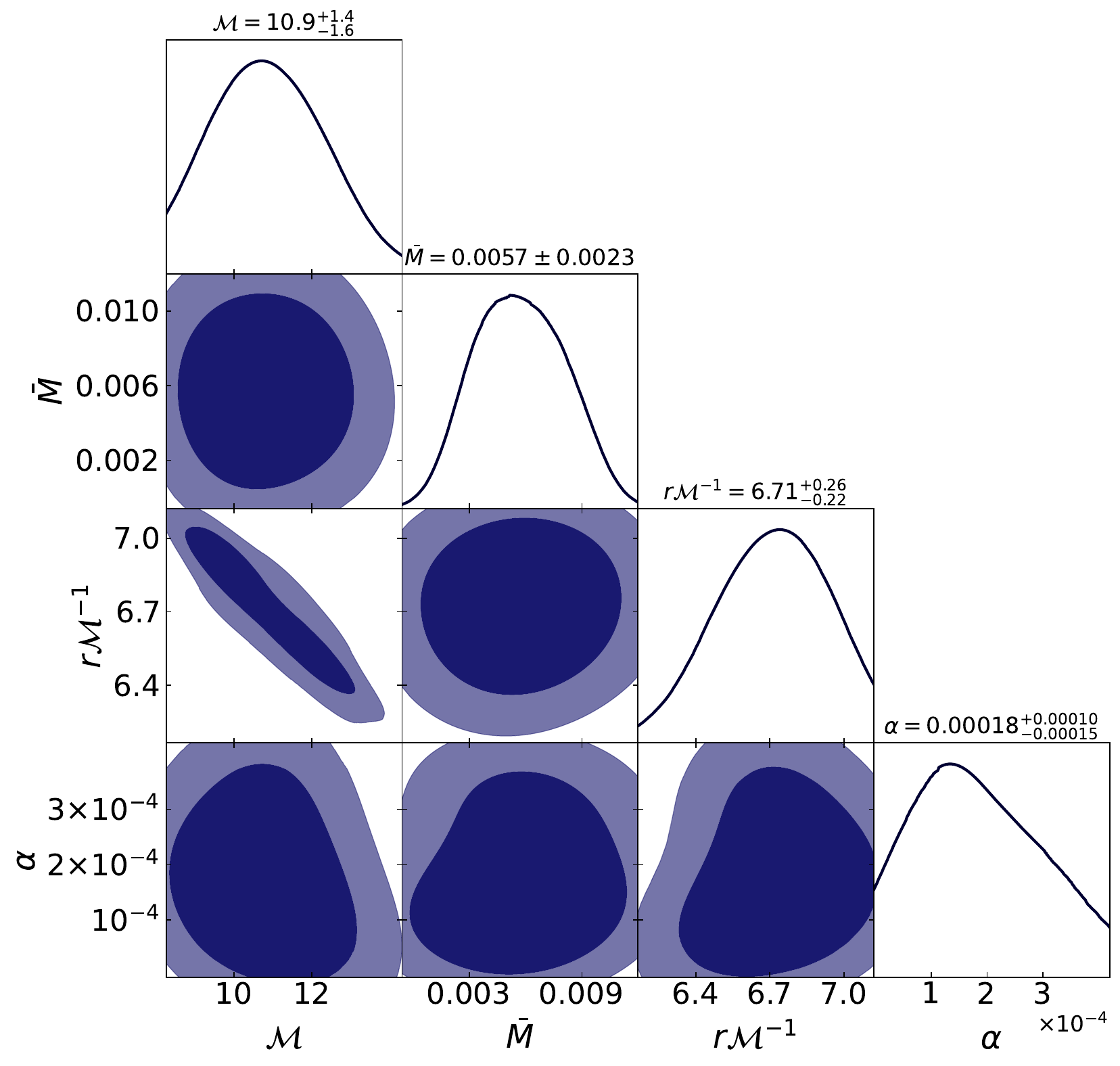}\hspace{0.3cm}
      \includegraphics[scale=0.285]{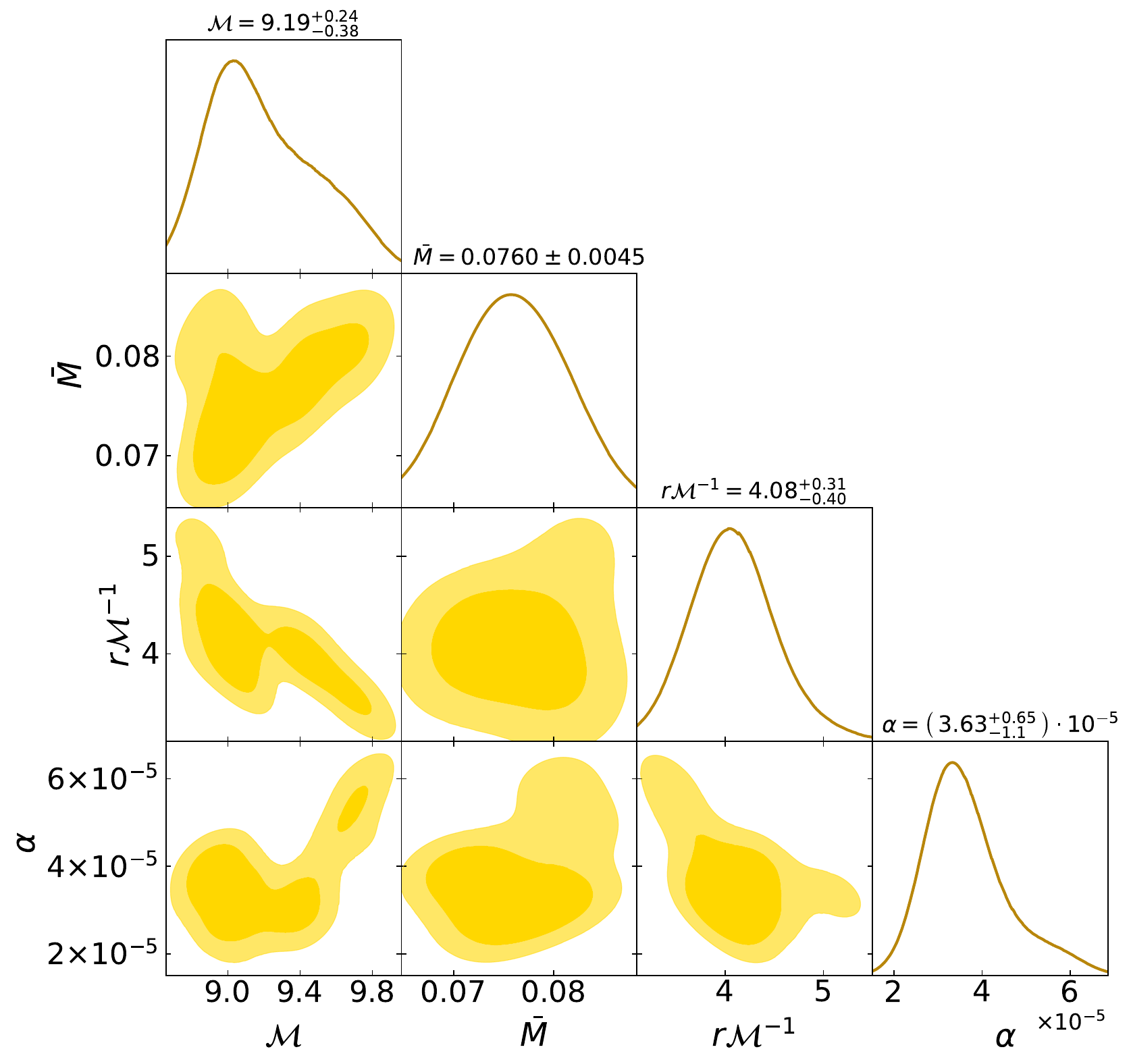}}
      \vspace{-0.2cm}
      \caption{MCMC outputs, showing the constraints on the parameters 
$\mathcal{M}$, $\bar{M}$, $r\mathcal{M}^{-1}$ and $\alpha$, using data from 
GRS 1915+105 (left) and H 1743–322 (right) BHs.}
\label{fig8}
\end{figure}
\begin{figure}[!h]
      \centerline{
     \includegraphics[scale=0.285]{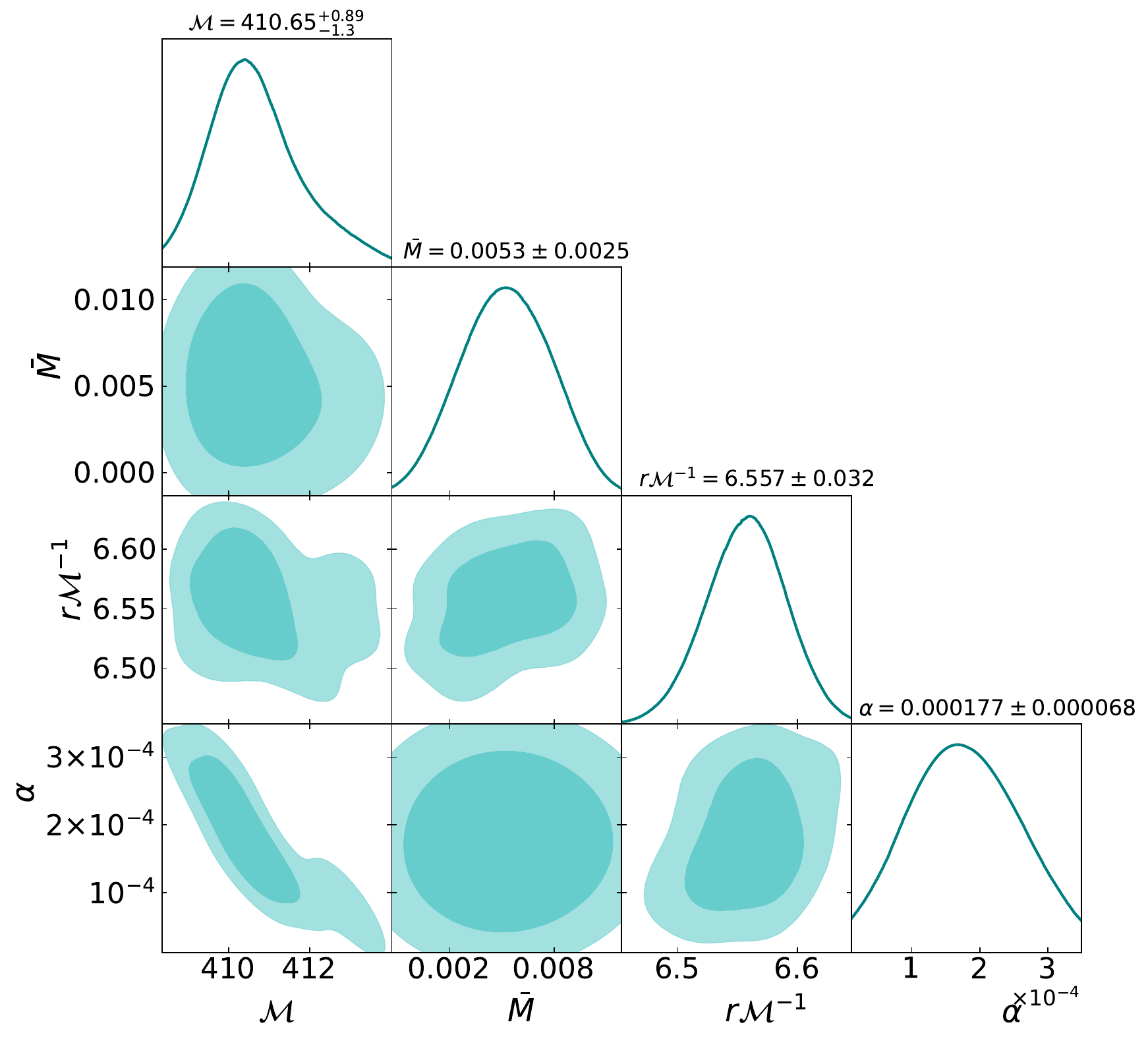}\hspace{0.3cm}
     \includegraphics[scale=0.285]{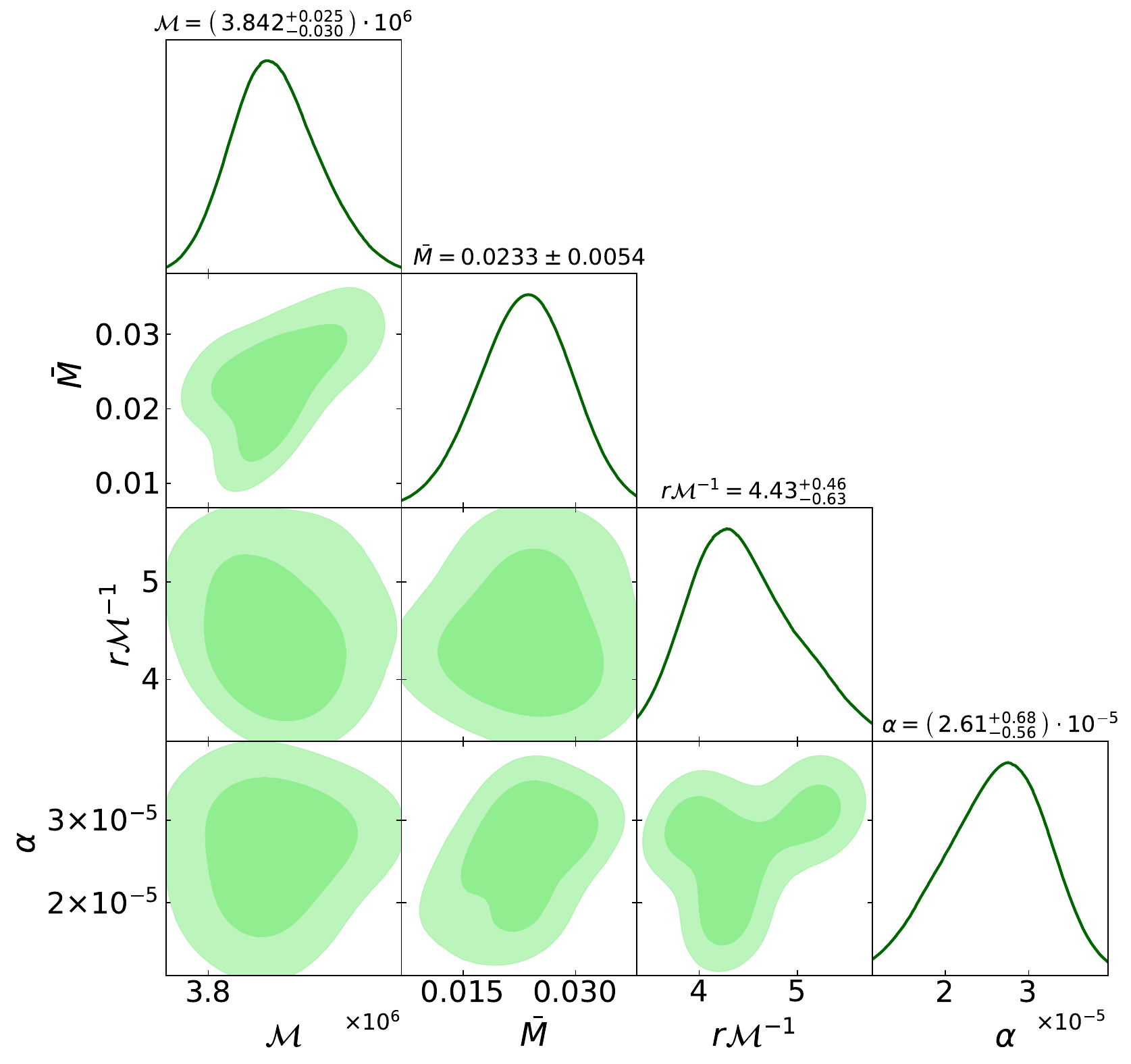}}
    \caption{MCMC outputs, showing the constraints on the parameters 
$\mathcal{M}$, $\bar{M}$, $r\mathcal{M}^{-1}$ and $\alpha$, using data from 
M82 X-1 (left) and Sgr A* (right) BHs.}
\label{fig9}
\end{figure}
Figs.~\ref{fig7} -- \ref{fig9} present the results of the MCMC analysis for 
six different BH systems. The left plot of Fig.~\ref{fig7} corresponds to the 
GRO J1655--40 system, using its measured mass along with the upper and lower 
QPO frequencies. Similarly, the right plot of this figure displays results for 
the XTE J1550--564 BH, based on the same set of observational data. The left 
plot Fig.~\ref{fig8} corresponds to the BH GRS 1915+105 and its right plot 
corresponds to the BH system H 1743+322. Similarly, the left plot of 
Fig.~\ref{fig9} is based on the data from the BH M82 X-1, and the right plot 
of it uses data from the BH Sgr A*. In this analysis, the contour plots 
illustrate the confidence intervals of the posterior probability distributions 
for all the parameters, showing levels of $68\%$ (1$\sigma$) and $95\%$ 
(2$\sigma$). The shaded areas within the contours represent these respective 
confidence regions. The dark shaded region is for $1\sigma$ confidence 
interval and the light shaded region is for $2\sigma$ confidence interval. 
Table \ref{tab2} shows the constraints on the model parameters for six 
different BHs. From Table \ref{tab2}, it can be observed that the 
constrained values of $\mathcal{M}$ are in good 
agreement with the available observational results, which shows the 
reliability of our analysis. For the parameter $\bar{M}$, the highest value is 
obtained for the BH system H 1743+322, while the lowest value corresponds to 
GRO J1655--40. In the case of the parameter $\alpha$, we find the maximum 
value for GRS~1915+105 and the minimum for Sgr~A*. These variations in the 
parameters indicate their source dependence. Finally, we calculated the 
theoretical values of the upper and lower QPO frequencies using the 
constrained parameters for all the BH systems listed in Table \ref{table1}, 
and the corresponding results are presented in Table \ref{tab3}. In 
Table~\ref{tab3}, we also present the deviations of the theoretically 
calculated QPO frequencies from the observed values, computed using the 
following formula:
\begin{equation}
\Delta \nu\, (\%) = \left| \frac{\nu_\text{ob} - \nu_\text{th}}{\nu_\text{ob}} \right| \times 100\%,
\end{equation}
where $\nu_\text{ob}$ and $\nu_\text{th}$ are the observed and theoretical 
values of QPO frequencies. In Table~\ref{tab3}, the 3rd column presents the 
deviations of the theoretical upper QPO frequencies from the observed values, 
while the 5th column shows the deviations of the theoretical lower QPO 
frequencies from the observed values. From this table, it is evident that the 
deviation between the observed and theoretical QPO frequencies is higher 
for stellar-mass and intermediate-mass BHs, whereas for the supermassive BH 
Sgr~A* the deviation is comparatively small.

\begin{table}[ht]
\centering
\caption{Estimated values of the parameters $\mathcal{M}$, $\bar{M}$, 
$r\mathcal{M}^{-1}$, and $\alpha$ derived from the MCMC analysis.}
\vspace{5pt}
\begin{tabular}{|c|c|c|c|c|}
\hline
\textbf{BH} &  $\mathcal{M}$ & $\bar{M}$ & $r\mathcal{M}^{-1}$ & $\alpha$ \\
\hline
GRO J1655--40 & $6.03 \pm 0.34$ & $0.0036 \pm 0.0018$ & $4.88^{+0.32}_{-0.40}$ & $(3.49^{+0.95}_{-1.1})\times 10^{-5}$\\
XTE J1550--564 & $9.29 \pm 0.86$ & $0.0048 \pm 0.0023$ & $6.15 \pm 0.13$ & $ (1.31 \pm 0.42)\times 10^{-4}$ \\
GRS 1915+105 & $10.9^{+1.4}_{-1.6}$ & $0.0057 \pm 0.0023$ & $6.71^{+0.26}_{-0.22}$ & $(1.80^{+0.10}_{-0.15}) \times 10^{-4}$ \\
H 1743+322 & $9.19^{+0.24}_{-0.38}$ & $0.0760 \pm 0.0045$ & $4.08^{+0.31}_{-0.40}$ & $(3.63^{+0.65}_{-1.1}) \times 10^{-5}$ \\
M82 X-1 & $410.65^{+0.89}_{-1.3}$ & $0.0053 \pm 0.0025$ & $6.557 \pm 0.032$ & $ (1.77 \pm 0.68)\times 10^{-4}$ \\
Sgr A* & $(3.842^{+0.025}_{-0.030}) \times 10^6$ & $0.0233 \pm 0.0054$ & $4.43^{+0.46}_{-0.63}$ & $(2.61^{+0.68}_{-0.56})\times 10^{-5}$ \\[3pt]
\hline
\end{tabular}
\label{tab2}
\end{table}

\begin{table}[!h]
\caption{Theoretical QPO frequencies as obtained from the RP model and 
their deviations from the corresponding observed values of chosen BH sources.}
\vspace{5pt}
\centering
\begin{tabular}{|c|c|c|c|c|}
\hline
\textbf{BH}  & \textbf{Upper Frequency (Hz)} & \textbf{$\Delta \nu_{U}$(\%)} & \textbf{Lower Frequency (Hz)} & \textbf{$\Delta \nu_{L}$ (\%)} \\
\hline
GRO J1655--40     & $497.87$ & $12.69$ & $317.26$ & $6.4$\\ \hline
XTE J1550--564   & $221.96$ & $10.13$ & $157.95$ & $14.15$\\ \hline
GRS 1915+105      & $198.71$ & $18.27$ & $127.97$ & $12.91$ \\
\hline
H 1743+322       & $209.29$ & $13.5$ & $134.35$ & $19.06$ \\ \hline
M82 X-1          & $5.57$ & $11.24$ & $3.96$ & $19.27$ \\ \hline
Sgr A*           & $1.33 \times 10^{-3}$ & $7.63$ & $0.81 \times 10^{-3}$ & $8.57$ \\
\hline
\end{tabular}  \label{tab3}
\end{table}

\section{Summery and Conclusion} \label{7}
The main goal of this work is to study the circular motion of a test particle 
around the BHs in the presence of a gravitational trace anomaly, viz., the GB 
trace anomaly and then to study the effect of this trace anomaly on the QPOs 
of BHs using different theoretical models. The findings of the study are 
summarised as follows.

In the first subsection of Section~\ref{3}, we examined the equations of 
motion for a test particle orbiting a BH under the influence of the GB trace 
anomaly. Fig.~\ref{fig1} presents the variation of the effective potential 
with the GB trace anomaly parameter~$\alpha$, showing that the peak of the 
potential increases as $\alpha$ grows. In the second subsection, we 
investigated the properties of circular orbits. Fig.~\ref{fig2} depicts the 
dependence of the angular momentum $\mathcal{J}$ and the specific energy 
$\mathcal{E}$ on $r\mathcal{M}^{-1}$ in the left and middle panels, 
respectively, while the right panel illustrates the variation of~$\mathcal{E}$ 
with $\mathcal{J}$. It is evident from Fig.~\ref{fig2} that the GB trace 
anomaly influence  the motion of particle compared to 
the GR case. In the third subsection, we analyzed the behavior of the innermost 
stable circular orbit as a function of~$\alpha$ for different values 
of~$\bar{M}$ as shown in Fig.~\ref{fig3}.

In Section~\ref{4} we studied the fundamental frequencies, i.e., Keplerian 
frequency and harmonic oscillation frequencies. In Subsection A, we calculated 
the Keplerian frequency $\Omega_{\phi}$ and studied the variation of 
$\Omega_{\phi}\mathcal{M}^{-1}$ with respect to $r\mathcal{M}^{-1}$. It is 
found that Keplerian frequency decreases with the radial distance and 
increases with the magnitude of the GB trace anomaly factor $\alpha$. The 
radial and vertical oscillation frequencies were calculated in Subsection~B. 

In Subsection~A of Section~\ref{5} we discussed the different QPO models, viz.,
PR, RP, WD, TD and ER2--ER4 models. In this subsection, Fig.~\ref{fig5} shows 
the relation between upper QPO frequency $\nu_U$ and lower QPO frequency 
$\nu_L$ calculated using those models, for different values of $\alpha$. This 
figure also includes reference lines for characteristic frequency ratios, with 
shaded regions between them marking distinct frequency ratio bands of 
twin-peak QPOs. It can be concluded from Fig.~\ref{fig5} that with higher 
values of $\alpha$ the relation between $\nu_U$ and $\nu_L$ deviates from the 
Schwarzschild case, highlighting the significant influence of the GB trace 
anomaly on QPO behaviour. In Subsection B, we have shown the 
dependence of QPO orbital radius with respect to $\alpha$ for frequency 
ratios 1:1, 3:2, 4:3 and 5:4 for all the above mentioned models along with 
the ISCO. It is clear from Fig.~\ref{fig6} that for the ER3 and TD models, the 
QPOs arise from significantly higher orbital radius. The ER2 and PR models also 
predict larger resonece radii. Moreover, the smallest QPO orbital is predicted 
by the RP model. It is to be noted that in all the models $r\mathcal{M}^{-1}$ 
increases with $\alpha$.

In Section~\ref{6}, we constrained the parameters of the considered BHs using 
an MCMC analysis. For this purpose, we employed mass and QPO data from six 
different BH systems, as listed in Table~\ref{table1}. The parameter priors 
were assumed to follow Gaussian distributions, as defined in Eq.~\eqref{eq43}. 
The log-likelihood functions for the upper and lower QPO frequencies are 
given in Eqs.~\eqref{eq45} and \eqref{eq46}, respectively. The results obtained 
from the MCMC analysis are summarized in Table~\ref{tab2}. As evident from 
the table, the constrained parameters also satisfy the condition in 
Eq.~\eqref{eq16}, which is a necessary requirement for the GB trace anomaly 
induced BHs. Furthermore, we calculated the theoretical values of the 
upper and lower QPO frequencies of the considered BHs for the RP QPO model 
using the parameter values obtained from the constraints. Our results show 
that the deviation between the observed and theoretical QPO frequencies is 
larger for the stellar-mass and intermediate-mass BHs, while for the 
supermassive BH Sgr~A*, the deviation is comparatively small.

In conclusion, our analysis demonstrates that the GB trace anomaly has a 
significant effect on both the geodesic motion of test particles and the QPO 
properties of BHs predicted by various theoretical models. The anomaly not 
only varies the effective potential and the location of the ISCO but also 
significantly modifies the correlations between upper and lower QPO 
frequencies, as well as the resonance radii in different models. The MCMC 
analysis, incorporating observational data from different BH systems, provides 
parameter constraints consistent with the theoretical requirements for GB 
trace anomaly induced BHs, thereby supporting the viability of this framework 
in explaining observed QPO behaviour. This work can be extened by incorporating 
more higher-order corrections to the spacetime metric and can aslo be extended 
to rotating BH solutions within GB trace anomaly gravity. 

\section*{Acknowledgements} UDG is thankful to the Inter-University Centre
for Astronomy and Astrophysics (IUCAA), Pune, India for the Visiting
Associateship of the institute.

\appendix*
\section{Expressions of $\mathcal{J}$, $\mathcal{E}$, $\Omega_r$, 
$\Omega_\theta$ and $\Omega_\phi$ } \label{App}
\vspace{-0.8cm}
\begin{equation}
{\mathcal{J}}^2 = \frac{
\begin{aligned}
& r^2 \Big\{
M^2 \Big(
38731 r^6 - 24720 r^5 r_{h} - 10950 r^4 r_{h}^2 - 93960 r^3 r_{h}^3 - 24015 r^2 r_{h}^4 - 22368 r r_{h}^5 + 65100 r_{h}^6
\Big) \alpha^2 \\
&\quad +  \bar{M}^2 \Big[
32400 M^4 r^6 r_{h}^4
- 1800 M^2 r^3 r_{h}^2 (23 r^3 - 24 r^2 r_{h} - 15 r r_{h}^2 - 24 r_{h}^3) \alpha \\
&\qquad + \Big(-6979 r^6 - 39120 r^5 r_{h} - 5850 r^4 r_{h}^2 - 19560 r^3 r_{h}^3 + 99435 r^2 r_{h}^4 + 53952 r r_{h}^5 + 31500 r_{h}^6\Big) \alpha^2
\Big]
\Big\}
\end{aligned}
}{
\begin{aligned}
& M^2 \Big(-116193 r^6 + 49440 r^5 r_{h} + 18250 r^4 r_{h}^2 + 140940 r^3 r_{h}^3 + 33621 r^2 r_{h}^4 + 29824 r r_{h}^5 - 83700 r_{h}^6\Big) \alpha^2 \\
&\quad + \bar{M}^2 \Big[
32400 M^4 r^6 (2r - 3r_{h}) r_{h}^3
+ 1800 M^2 r^3 r_{h}^2 (69 r^3 - 48 r^2 r_{h} - 25 r r_{h}^2 - 36 r_{h}^3) \alpha \\
&\qquad + \Big(20937 r^6 + 78240 r^5 r_{h} + 9750 r^4 r_{h}^2 + 29340 r^3 r_{h}^3 - 139209 r^2 r_{h}^4 - 71936 r r_{h}^5 - 40500 r_{h}^6\Big) \alpha^2
\Big]
\end{aligned}
}
\end{equation}
\begin{equation}
\begin{aligned}
{\mathcal{E}}^2 = &\left(1 - \frac{r_{h}}{r}\right)
\Bigg( 1 + \frac{(23 r^2 + 11 r r_{h} + 6 r_{h}^2) \alpha}{18 M^2 r^3 r_{h}} \\
&\quad - \frac{
e^{-2\phi} M^2 \left(38731 r^5 + 26371 r^4 r_{h} + 22721 r^3 r_{h}^2 - 769 r^2 r_{h}^3 - 5572 r r_{h}^4 - 9300 r_{h}^5\right)
}{32400 \bar{M}^2 M^4 r^6 r_{h}^3} \alpha^2
 \\
&\quad - \frac{\bar{M}^2 \left(6979 r^5 + 26539 r^4 r_{h} + 28489 r^3 r_{h}^2 + 33379 r^2 r_{h}^3 + 13492 r r_{h}^4 + 4500 r_{h}^5\right)}
{32400 \bar{M}^2 M^4 r^6 r_{h}^3} \alpha^2 \Bigg) \\
&\times
\left(
\frac{
\begin{aligned}
& M^2 \left(38731 r^6 - 24720 r^5 r_{h} - 10950 r^4 r_{h}^2 - 93960 r^3 r_{h}^3 - 24015 r^2 r_{h}^4 - 22368 r r_{h}^5 + 65100 r_{h}^6\right) \alpha^2 \\
&\quad + e^{2\phi} \bar{M}^2 \Big[
32400 M^4 r^6 r_{h}^4
- 1800 M^2 r^3 r_{h}^2 (23 r^3 - 24 r^2 r_{h} - 15 r r_{h}^2 - 24 r_{h}^3) \alpha \\
&\qquad + \left(-6979 r^6 - 39120 r^5 r_{h} - 5850 r^4 r_{h}^2 - 19560 r^3 r_{h}^3 + 99435 r^2 r_{h}^4 + 53952 r r_{h}^5 + 31500 r_{h}^6\right) \alpha^2
\Big]
\end{aligned}
}{
\begin{aligned}
& M^2 \left(-116193 r^6 + 49440 r^5 r_{h} + 18250 r^4 r_{h}^2 + 140940 r^3 r_{h}^3 + 33621 r^2 r_{h}^4 + 29824 r r_{h}^5 - 83700 r_{h}^6\right) \alpha^2 \\
&\quad + e^{2\phi} \bar{M}^2 \Big[
32400 M^4 r^6 (2r - 3r_{h}) r_{h}^3
+ 1800 M^2 r^3 r_{h}^2 (69 r^3 - 48 r^2 r_{h} - 25 r r_{h}^2 - 36 r_{h}^3) \alpha \\
&\qquad + \left(20937 r^6 + 78240 r^5 r_{h} + 9750 r^4 r_{h}^2 + 29340 r^3 r_{h}^3 - 139209 r^2 r_{h}^4 - 71936 r r_{h}^5 - 40500 r_{h}^6\right) \alpha^2
\Big]
\end{aligned}
}
\right)
\end{aligned}
\end{equation}

\begin{equation}
\Omega_r = 
\sqrt{ \frac{(2 \mathcal{M} - r)}{518400\, \mathcal{M}^3 M^4 r^8 \bar{M}^2} \left(
\begin{aligned}
  &144000 \alpha^2 \mathcal{M}^5 \bar{M}^2 
  + 172800 \alpha \mathcal{M}^4 M^2 r^3 \bar{M}^2 
  + 215872 \alpha^2 \mathcal{M}^4 r \bar{M}^2 
  + 259200 \mathcal{M}^3 M^4 r^6 \bar{M}^2 \\
  &+ 158400 \alpha \mathcal{M}^3 M^2 r^4 \bar{M}^2 
  + 267032 \alpha^2 \mathcal{M}^3 r^2 \bar{M}^2 
  + 165600 \alpha \mathcal{M}^2 M^2 r^5 \bar{M}^2 
  + 113956 \alpha^2 \mathcal{M}^2 r^3 \bar{M}^2 \\
  &+ 53078 \alpha^2 \mathcal{M} r^4 \bar{M}^2 
  + 6979 \alpha^2 r^5 \bar{M}^2 
  + 297600 \alpha^2 \mathcal{M}^5 M^2 
  + 89152 \alpha^2 \mathcal{M}^4 M^2 r \\
  &+ 6152 \alpha^2 \mathcal{M}^3 M^2 r^2 
  - 90884 \alpha^2 \mathcal{M}^2 M^2 r^3 
  - 52742 \alpha^2 \mathcal{M} M^2 r^4 
  - 38731 \alpha^2 M^2 r^5 
\end{aligned}
\right) } \label{Omega}
\end{equation}

\begin{equation}
\Omega_{\theta} = \Omega_{\phi} = 
\frac{1}{720} \sqrt{
\frac{
\begin{aligned}
&2016000\, \alpha^2 \mathcal{M}^6 \bar{M}^2 
+ 1382400\, \alpha\, \mathcal{M}^5 M^2 r^3 \bar{M}^2 
+ 1726464\, \alpha^2 \mathcal{M}^5 r\, \bar{M}^2 \\
&+ 518400\, \mathcal{M}^4 M^4 r^6 \bar{M}^2 
+ 432000\, \alpha\, \mathcal{M}^4 M^2 r^4 \bar{M}^2 
+ 1590960\, \alpha^2 \mathcal{M}^4 r^2 \bar{M}^2 \\
&+ 345600\, \alpha\, \mathcal{M}^3 M^2 r^5 \bar{M}^2
- 156480\, \alpha^2 \mathcal{M}^3 r^3 \bar{M}^2 \\
&- 165600\, \alpha\, \mathcal{M}^2 M^2 r^6 \bar{M}^2 
- 23400\, \alpha^2 \mathcal{M}^2 r^4 \bar{M}^2 
- 78240\, \alpha^2 \mathcal{M}\, r^5 \bar{M}^2 \\
&- 6979\, \alpha^2 r^6 \bar{M}^2 
+ 4166400\, \alpha^2 \mathcal{M}^6 M^2 
- 715776\, \alpha^2 \mathcal{M}^5 M^2 r \\
&- 384240\, \alpha^2 \mathcal{M}^4 M^2 r^2 
- 751680\, \alpha^2 \mathcal{M}^3 M^2 r^3 
- 43800\, \alpha^2 \mathcal{M}^2 M^2 r^4 \\
&- 49440\, \alpha^2 \mathcal{M} M^2 r^5 
+ 38731\, \alpha^2 M^2 r^6
\end{aligned}
}{
\mathcal{M}^3 M^4 r^9 \bar{M}^2
}
}
\label{omegaphi}
\end{equation}


\end{document}